\documentstyle[aps,epsfig]{revtex} 
\newcommand{\be}{\begin{equation}}
\newcommand{\ee}{\end{equation}\noindent}
\newcommand{\eei}{\end{equation}}
\newcommand{\bea}{\begin{eqnarray}}
\newcommand{\eea}{\end{eqnarray}\noindent}
\newcommand{\eeai}{\end{eqnarray}}

\newcommand{\hf} {{1\over2}}
\newcommand{\nonu}{\nonumber\\}
\def\eq#1{(\ref{#1})}
\begin{document}
\draft
\title{Renormalization of composite operators}
\author{J. Polonyi$^{1,2}$ and K. Sailer$^3$}
\address{$^1$ Institute for Theoretical Physics, Louis Pasteur University,
Strasbourg, France}
\address{$^2$ Department of Atomic Physics, Lorand E\"otv\"os University, 
Budapest, Hungary}
\address{$^3$ Department for Theoretical Physics, University of  
Debrecen, Hungary}
\date{\today}
\maketitle
\begin{abstract}
The blocked composite operators are defined in the one-component
Euclidean scalar field theory, and shown to generate a linear
transformation of the operators, the operator mixing. This
transformation allows us to introduce the parallel transport
of the operators along the RG trajectory. The connection on
this one-dimensional manifold governs the scale evolution of
the operator mixing. It is shown that the solution of the
eigenvalue problem of the connection gives the various scaling regimes
and the relevant operators there. The relation to perturbative
renormalization is also discussed in the framework of the $\phi^3$
theory in dimension $d=6$.
\end{abstract}
\pacs{11.15Tk, 11.10Gh}

\renewcommand{\thefootnote}{\fnsymbol{footnote}}
\def\thefootnote{\arabic{footnote}}
\setcounter{footnote}{1}
\def\theequation{\arabic{section}.\arabic{equation}}

\normalsize
\setcounter{table}{0}

\section{Introduction}\label{Introduc}
The RG strategy is to follow the evolution of the coupling constants
in the observational scale in order to identify the important
interactions at different scales. This is achieved by the blocking,
the successive lowering of the
UV cut-off and the tracing of the resulting blocked, renormalized
action \cite{Wil74,Weg73}. 
There is an alternative method, implicit in this procedure, 
where the original cut-off and action are kept but the operators
are modified to cover less modes while keeping the expectation
values fixed. This paper outlines this latter method in
a more systematic manner and compares it with the traditional one.

A similar question arises in the renormalization of composite
operators where the goal is to render  the Green's functions
containing the insertion of some local operators which are not
in the action finite as the cut-off is removed. The perturbative
treatment of this problem proceeds by the introduction of additional
counterterms in the action in such a manner that the original
Green's functions remain unchanged but those with the insertion of the
composite operators turn  out to be finite \cite{Wei96,Pok,Zin89,Col84}.

This rather complicated procedure attempts to perform an amazing
project, the renormalization of an
otherwise non-renormalizable model. This happens because the 
insertion of the non-renormalizable composite operators in the Green's
functions can be achieved by introducing them in the action with a source
term and taking the derivative of the logarithm of the partition
function with respect to their source. As long as the Green's functions
are finite so is the partition function containing
non-renormalizable operators! The resolution of this apparent
paradox is that the rendering of more Green's functions with the
composite operator insertion finite requires more counterterms.
At the end we should need infinitely many counterterms to complete
the project, just as when we would attempt to remove the cut-off
in a non-renormalizable model. 

In the traditional multiplicative renormalization of quantum field
theory the irrelevant, non-renormalizable coupling constants
are ignored. In fact, the derivation of the renormalization group
equations in renormalizable quantum field theory is based on the
compensation of the change of the scale by the adjusting of the relevant
(or marginal) coupling constants. This procedure
is made insensitive to the irrelevant terms by
considering the asymptotic scaling regime only and by ignoring the
vanishing contributions as the cut-off is removed. It cannot
help us to understand crossovers or the competition of different
fixed points, in general. One might object that irrelevant operators
do appear in the renormalization of composite
operators within the traditional multiplicative renormalization scheme.
But their evolution is determined by the requirement that the Feynman
graphs considered remain finite. This prescription misses the
finite, physical part of the renormalization.
The blocking procedure, followed in the present paper provides
a renormalization scheme where the evolution of all coupling
constants is well defined and we are not confined into the
asymptotic scaling regimes.

The non-perturbative renormalization due to blocking has proven
to be effective in establishing renormalizability, as well.
The old-fashioned perturbative proof of renormalizability which
is made involved by the tracing of overlapping divergences
\cite{Col84} is simplified enormously by considering a
non-perturbative renormalization scheme \cite{Polch84,Hugh88}. In fact,
the renormalizability turns out to be the absence of the UV Landau
pole. This pole can easily be avoided in the framework of the
loop expansion for asymptotically free models.

Our technique to follow the mixing of composite operators by adding
them to the action has already been used in the framework of the
perturbative, multiplicative renormalization group scheme.
In the perturbative studies of the renormalizability of Green's 
functions one or more composite operators were inserted by 
coupling them to spacetime-dependent external sources and adding the
resulting expressions to the action. Then, the functional differentiation of 
the generating functional of the connected Green's functions was performed 
w.r.t. these sources \cite{Hugh88,Shore91,Kell92}.
It was also shown that taking the identity operator into account
explicitly in the action has a particular importance for the renormalization 
of
multilocal composite operators \cite{Becchi96}.

Recently, there have been made attempts to give a covariant geometric
interpretation to the RG flow \cite{Dolan95a}-\cite{Dolan99}. The space of
coupling constants is considered as a differentiable manifold and the RG
flow, expressed by means of Lie transport, is interpreted as a
particular mapping on it. It has been established by introducing a
metric in the coupling constant space that the RG flow is not
geodesic in general \cite{Dolan97,Brody98}. Also the definition of
the connection appears problematic \cite{Dolan95a}, although recently
a non-metric-compatible definition was given \cite{Dolan99}.

Our geometric interpretation is less ambitious. We try to implement
differential geometric notions along a one-dimensional manifold only,
the RG-trajectory. It is based on parallel transport along the RG
trajectory, introduced differently as generally used in the literature.
Namely, the space of operators (symbols) is considered as a tangent
space at each point of the RG trajectory. This has the advantage that
no metric is needed and the connection on the RG trajectory arises
in a rather straightforward way, cf. Eq. \eq{covcon} below.
It is the geometry for the RG trajectory only which is covered in
this manner, ignoring the geometrical aspects of the more
complicated space parametrised by the coupling constants.

Nevertheless, our method reveals some advantageous features.
The local properties of the RG trajectories, given by the connection,
are determined in a unique manner at any point in the coupling constant
space. It will be shown that the connection can directly be obtained
from the beta-functions in terms of the blocked coupling constants.
The usual blocking is based on local quantities of the RG flow, such as
the beta-functions which are sufficient for the local studies in the vicinity
of a given fixed point. But the RG flow of more realistic models visits
several scaling regimes between the UV and the IR regimes and the
determination of the set of important parameters may require global
methods which can take into account the interplay between different
scaling regimes \cite{Alex}. The operator renormalization includes in a
natural manner a global quantity, called sensitivity matrix, needed
for such analysis. A flow equation will be derived for the sensitivity
matrix, expressing that the evolution is governed by the connection.
This flow equation enables us to perform the renormalization of
composite operators beyond the perturbation expansion. In this manner
the operator renormalization is better suited for the studies of
models with non-trivial IR scaling law than the traditional blocking.

In Sect. \ref{Toymodel} a toy model is presented explaining the origin
of operator mixing. Then, the idea of blocking is extended to
operators in one-component scalar field theory in Sect. \ref{Operamix}
and in Sect. \ref{Parh} the differential geometric meaning of blocking
the operators is
clarified. The possibility of finding the scaling operators at an
arbitrary scale by means of solving the eigenvalue problem of the
sensitivity matrix is discussed in Sect. \ref{Classific}.
The construction of the blocked operator and the renormalized
perturbation expansion (RPE) are compared in Sect. \ref{Compar}.
The resolution of the paradoxon of renormalizing irrelevant composite
operators in a renormalizable theory is also discussed.
The operator mixing matrix is determined for $\phi^3$ theory in
dimension $d=6$ in a restricted operator basis  in the
independent mode approximation (IMA) in two different ways
in Sects. \ref{indiropmx} and \ref{diropmx}. The agreement of these
results with the one-loop perturbative ones are shown in Sect.
\ref{Opmper}. Finally, the main results of the paper are summarized in
Sect. \ref{summary}.

\section{Toy model for operator mixing}\label{Toymodel}

\setcounter{equation}{0}

Let us consider a zero-dimensional model with two degrees of freedom,
$x$ (`the low-frequency one') and
$y$ (`the high-frequency one'), and the bare action
\begin{eqnarray}
S(x,y) ={1\over2}s_xx^2+{1\over2}s_yy^2 
+\sum_{n=0}^\infty g_n (x+y)^n
\end{eqnarray} 
with the bare couplings $g_n$. We are looking for the blocked action $S(x)$
obtained by integrating out the degree of freedom $y$,
\begin{eqnarray}
\label{Spx}
  e^{-S (x) } &=& \int dy e^{- S(x,y)}.
\end{eqnarray} 
The composite operators of the bare theory are defined as $(x+y)^n$ with 
$n=0,1,2,\ldots$ According to  relation (\ref{Spx}), the blocked action $S(x)$ as the 
function of the bare coupling constants $g_n$ can be considered as the
generator function for the composite operators for a given value $x$, 
\begin{eqnarray}
\label{State1toy}
   \frac{ \partial S (x)}{\partial g_n} &=&
   \frac{ \int dy (x+y)^n e^{- S(x,y)}  }{
          \int dy e^{- S(x,y)  }         },
\end{eqnarray}
i.e. the partial derivative of the blocked action w.r.t. one of the
bare couplings $g_n$ is equal to the `high-frequency' average of the
corresponding bare composite operator $(x+y)^n$. As a consequence the relation
\begin{eqnarray}
\label{expv1}
\frac{\int dx\frac{\partial S(x)}{\partial g_n}e^{-S(x)}}{\int dxe^{-S(x)}} 
&=&\frac{\int dxdy(x+y)^ne^{-S(x,y)}}{\int dxdye^{-S(x,y)}}
\end{eqnarray}
holds, i.e. the partial derivative of the blocked action w.r.t. the
bare coupling $g_n$ can be interpreted as the blocked operator that
provides the same expectation value in the blocked theory as the
bare operator $(x+y)^n$ does in the bare theory. Let us
introduce the notation $\{ x^n \} = \partial S (x) / \partial
g_n$ for it. The action $S (x)$ can be expanded in the terms of the
base operators as 
\begin{eqnarray}
S (x) ={1\over2}s_xx^2+ \sum_{m=0}^\infty g_m'  x^m
\end{eqnarray}
with the blocked couplings $g_m'  = g_m' (g_0,g_1,g_2,\ldots)$
$(m=0,1,2, \ldots )$ being functions of the bare couplings. 
Using this expansion, the
blocked operators can be rewritten in the form:
\begin{eqnarray}
\label{opmix}
  \{ x^n \}&=& \frac{\partial S (x) }{ \partial g_n } 
       = \sum_{m=0}^\infty  x^m S_{mn}
\end{eqnarray}
with the help of the matrix 
\begin{eqnarray}
\label{mixmat}
   S_{mn} &=& \frac{\partial g_m ' }{ \partial g_n} .
\end{eqnarray} 
Eq. (\ref{opmix}) has the simple interpretation that the blocked
operator $\{ x^n \}$  obtained from the bare operator $(x+y)^n$
 by integrating out the degree of
freedom $y$ is a linear combination of the base operators $x^m$
with the coefficients given by the operator mixing matrix $S_{mn}$.
Thus, the operator with  which one can reproduce the vacuum expectation
value of a bare base operator in the blocked theory is the linear combination
of all the base operators with arguments projected into the subspace
of the degree of freedom $x$ left over.

Above we chose base operators in the bare theory and searched for
operators in the blocked theory that reproduce their vacuum expectation
values. Also the opposite question can be formulated: we
choose a basis of operators in the blocked theory, $x^n$ with
$n=0,1,2,\ldots$ and search for the operators $\lbrack (x+y)^n \rbrack$ 
of the bare theory that reproduce their expectation values,
i.e. for which
\begin{eqnarray}
\label{expv2}
\frac{\int dxdy\lbrack(x+y)^n\rbrack e^{-S(x,y)}}{\int dxdye^{-S(x,y)}}
&=&\frac{\int dxx^ne^{-S(x)}}{\int dxe^{-S(x)}} 
\end{eqnarray}
assuming that the bare theory underlying the blocked one is known.
Looking for the bare operators in the general form
$ \lbrack (x+y)^n \rbrack= \sum_{m=0}^\infty \alpha_m^{(n)} 
(x+y)^m $, and using Eqs. (\ref{expv1}) and (\ref{opmix}),
Eq. (\ref{expv2}) leads to the relation 
\begin{eqnarray}
   \sum_{m=0}^\infty x^{n'} S_{n'm} \alpha_m^{(n)}&=&x^n
\end{eqnarray}
and one finds that the matrix $\alpha_m^{(n)} = (S^{-1} )_{mn}$
is the inverse of the operator mixing matrix given by
Eq. (\ref{opmix}).
Thus, the bare operators looked for are given by
\begin{eqnarray}
\label{invblop}
\lbrack (x+y)^n\rbrack &=&\sum_{m=0}^\infty (x+y)^m(S^{-1} )_{mn}.
\end{eqnarray}
We see again, that the bare operators reproducing the expectation
values of given base operators of the blocked theory in the bare one
are the linear combinations of the base operators.

It is instructive to generalize the toy model for three variables,
\begin{eqnarray}
S(x,y,z) ={1\over2}s_xx^2+{1\over2}s_yy^2+{1\over2}s_zz^2 
+\sum_{n=0}^\infty g_n (x+y+z)^n.
\end{eqnarray} 
The blocking gives rise the chain of effective theories,
\bea
e^{-S(x,y)}=\int dze^{-S(x,y,z)},& \qquad&
e^{-S(x)}=\int dye^{-S(x,y)},
\eea
and operator mixing
\bea
\{x^n\}_z={\partial S(x,y,z)\over\partial g_n}=(x+y+z)^n, &\qquad&
\{x^n\}_y={\partial S(x,y)\over\partial g_n}=
-{{\partial\over\partial g_n}\int dze^{-S(x,y,z)}\over\int
  dze^{-S(x,y,z)}} ,
\nonumber\\
\{x^n\}_x=&{\partial S(x)\over\partial g_n}&=
-{{\partial\over\partial g_n}\int dydze^{-S(x,y,z)}\over
\int dydze^{-S(x,y,z)}}.
\eea
These relations yield
\bea
\{x^n\}_y={\int dz\{x^n\}_ze^{-S(x,y,z)}\over\int dze^{-S(x,y,z)}}, &\qquad&
\{x^n\}_x={\int dy\{x^n\}_ye^{-S(x,y)}\over\int dye^{-S(x,y)}},
\eea
indicating that the evolution of the operators comes from the elimination
of the field variable in their definition. One can compute
the expectation value of $\{x^n\}$ at any level,
\be
{\int dxdydz\{x^n\}_ze^{-S(x,y,z)}\over\int dxdydze^{-S(x,y,z)}}
={\int dxdy\{x^n\}_ye^{-S(x,y)}\over\int dxdye^{-S(x,y)}}
={\int dx\{x^n\}_xe^{-S(x)}\over\int dxe^{-S(x)}}.
\ee
In other words, the cut-off dependence of the renormalized operators is 
introduced in such a manner that the expectation values can be 
recovered for an arbitrary value of the cut-off.

We learned on the above discussed toy-model that operator mixing
is an immediate consequence  of  keeping the expectation values
unchanged under integrating out degrees of freedom.

\section{Operator mixing}\label{Operamix}

\setcounter{equation}{0}

Here we define blocked operators in quantum field theory following the
line illustrated by the zero-dimensional  toy-model of the previous
section.

Let us consider the  theory given by the bare action
\begin{eqnarray}
\label{bareact}
    S_\Lambda[\phi] &=& \int dx \sum_n G_n (x, \Lambda ) O_n(\phi(x)) ,
\end{eqnarray}
where $O_n (\phi (x)) $ represents a complete set of local operators
(function of $\phi(x)$ and its space-time derivatives)
coupled to the external sources $G_n(x,\Lambda)$.
The scalar field $\phi(x)=\varphi_k (x) + \chi_k (x)$
is decomposed into a low-frequency part 
\be
\varphi_k(x)=\sum_{|p|\le k}\phi_pe^{ipx}
\ee
and a high-frequency part 
\be
\chi_k(x)=\sum_{|p|\in\lbrack k,\Lambda\rbrack}\phi_pe^{ipx}
\ee
with the UV cut-off $\Lambda$ and the sharp moving
 cut-off $k$. Here inhomogeneous external sources have been introduced for
 later convenience and assumed that their zero modes $g_n (\Lambda )$ are
 separated, 
\be
G_n(x,\Lambda)=g_n(x,\Lambda)+g_n(\Lambda).
\ee 

Notice that the variables $n$ and $x$
of the operator $ O_n(\phi(x))$ identify a member of the complete set.
One can simplify the expressions by introducing a single index $\tilde n$
for the pair $(n,x)$ and $\sum_{\tilde n}$ for the integral and sum
$\sum_n\int dx$, e.g.
\be
S_\Lambda[\phi]=\sum_{\tilde n}G_{\tilde n}(\Lambda )O_{\tilde n}(\phi).
\ee

It is required that the partition function of the blocked theory $Z_k$ and that
of the bare one $Z_\Lambda$ are identical:
\be\label{baregen}
\frac{ \int {\cal D} \varphi_k {\cal D} \chi_k
 e^{- S_\Lambda \lbrack \varphi_k + \chi_k \rbrack } 
         }{
    \left.    \int {\cal D} \varphi_k {\cal D} \chi_k
  e^{- S_\Lambda \lbrack \varphi_k + \chi_k \rbrack } 
     \right|_{g_n (x, \Lambda) =0}}
=\frac{ \int {\cal D} \varphi_k
 e^{- S_k \lbrack \varphi_k  \rbrack }}{
    \left.    \int {\cal D} \varphi_k 
  e^{- S_k \lbrack \varphi_k  \rbrack } 
     \right|_{g_n (x,k) =0}}
\ee
that leads to the definition of the blocked action $S_k\lbrack\varphi_k\rbrack$
up to a constant,
\begin{eqnarray}
\label{blactdef}
   e^{ - S_k \lbrack \varphi_k \rbrack }
      &=&
  \int {\cal D} \chi_k  e^{ - S_\Lambda  \lbrack \varphi_k + \chi_k
    \rbrack } .
\end{eqnarray}

As shown in  \cite{Weg73} the blocked action satisfies the
Wegner-Houghton equation
\begin{eqnarray}
\label{WHeq}
    \frac{\partial S_k \lbrack \varphi_k \rbrack }{
        \partial k} 
             &=&
    -\lim_{\delta k\to0} {1\over2\delta k}Tr\ln
{\delta^2S_k\over\delta\varphi_k\delta\varphi_k}
\end{eqnarray}
where the trace is taken over the functional space with Fourier modes
$k-\delta k<p<k$. The blocked action can be expanded in the base operators as
\begin{eqnarray}
\label{blact}
S_k\lbrack\varphi_k\rbrack&=&\sum_{\tilde n}G_{\tilde n}(k)
O_{\tilde n}(\varphi_k) 
\end{eqnarray}
and (\ref{WHeq}) rewritten in the form of a coupled set of
differential equations
\begin{eqnarray}
\label{couWHeq}
    k \partial_k G_{\tilde n} (k) &=& \beta_{\tilde n}(k,G)
\end{eqnarray}
 for the blocked external sources 
 $ G_{\tilde n} (k)$. The right hand sides of these equations represent
 explicit expressions for the beta-functions as functions of the
 blocked sources. 

Let us now turn to the definition of blocked operators. From
Eq. (\ref{blactdef}), one gets after functional differentiation
\begin{eqnarray}
\label{State1}
\frac{ \delta S_k \lbrack \varphi_k \rbrack }{ \delta G_{\tilde n}(\Lambda)  }
&=&\frac{ \int {\cal D} \chi_k O_{\tilde n} (\varphi_k+\chi_k)
          e^{  - S_\Lambda  \lbrack \varphi_k + \chi_k
    \rbrack }       }{
           \int {\cal D} \chi_k   
 e^{- S_\Lambda  \lbrack \varphi_k + \chi_k   \rbrack }     
                     } ,
\end{eqnarray}
i.e.  the functional derivative of the blocked action w.r.t. any of
the bare sources $G_{\tilde n} ( \Lambda )$ is equal to the
high-frequency average of the
corresponding bare operator $O_{\tilde n}(\varphi_k +\chi_k)$.
Thus the functional derivative  
$   \delta S_k \lbrack \varphi_k \rbrack /$
 $ \delta G_{\tilde n}(\Lambda)$ reproduces the expectation value of the bare
operator $ O_{\tilde n}(\varphi_k+\chi_k)$,
\be
\label{expval1}
\frac{ \int {\cal D} \varphi_k 
{\delta S_k \lbrack \varphi_k \rbrack\over\delta G_{\tilde n}(\Lambda)}
     e^{ - S_k \lbrack \varphi_k \rbrack }
         }{
 \int {\cal D} \varphi_k   e^{ - S_k \lbrack \varphi_k \rbrack}}
=\frac{ \int {\cal D} \varphi_k {\cal D} \chi_k 
        O_{\tilde n}(\varphi_k+\chi_k)
  e^{- S_\Lambda  \lbrack \varphi_k + \chi_k   \rbrack }  
           }{
 \int {\cal D} \varphi_k {\cal D} \chi_k 
  e^{- S_\Lambda  \lbrack \varphi_k + \chi_k   \rbrack }  
            } .
\ee

As a result the functional derivative of the blocked action w.r.t.
one of the bare external sources $ G_{\tilde n} (\Lambda)$ 
can be interpreted as the operator obtained by blocking from the base
operator $O_{\tilde n}(\varphi_k+\chi_k)$. Let us introduce the notation
\begin{eqnarray}
\label{blopenot}
\{ O_{\tilde n}(\varphi_k)\}_k=\frac{\delta S_k\lbrack\varphi_k\rbrack }{
\delta G_{\tilde n}(\Lambda)}
\end{eqnarray}
for the corresponding blocked operator.  Making use of  expansion
(\ref{blact}) of
the blocked action, the operator mixing
\begin{eqnarray}
\label{blopexp}
  \{ O_{\tilde n}(\varphi_k)\}_k &=&
     \sum_{\tilde m}\frac{ \delta G_{\tilde m}(k)}{
              \delta G_{\tilde n} (\Lambda )  }
         \frac{ \delta S_k \lbrack \varphi_k \rbrack }{
                \delta  G_{\tilde m}(k) }
= \sum_{\tilde m} O_{\tilde m}(\varphi_k)S_{\tilde m\tilde n}(k,\Lambda)
\end{eqnarray}
can be given in terms of the sensitivity matrix 
\begin{eqnarray}
\label{mixmatdf}
S_{\tilde m\tilde n} (k,\Lambda)&=&\frac{ \delta G_{\tilde m}(k)}
{\delta G_{\tilde n}(\Lambda)},
\end{eqnarray}
which shows the dependence of the blocked coupling
constants on the initial values of the RG flow. As such it is a global
feature of the RG trajectory.

Conversely, if one chooses the base operators $O_{\tilde n}(\varphi_k)$
at the scale $k$ then one can search for the bare operator 
$\lbrack O_{\tilde n}(\varphi_k+\chi_k)\rbrack_k$ that reproduces the
vacuum expectation value of $O_{\tilde n}(\varphi_k)$. An
argument similar to that for the toy-model in the previous section leads to
\begin{eqnarray}\label{opkevsz}
\lbrack O_{\tilde n}(\varphi_k+\chi_k)\rbrack_k  
&=&\sum_{\tilde m}O_{\tilde m}(\varphi_k+\chi_k)
(S^{-1}(k,\Lambda))_{\tilde m\tilde n}.
\end{eqnarray}
Now we pass the RG trajectory in the opposite direction (from small
$k$ to large $k$ values) when asking for the bare operator that reproduces
the same vacuum expectation value as a given operator at the scale of
small $k$. Therefore, the answer is given in terms of the inverse
of the matrix $S_{\tilde m\tilde n} (k,\Lambda)$ describing the mixing of
operators if the RG trajectory is passed from large $k$ values  towards
the small ones as it happened when defining the blocked operators
$\{ O_{\tilde n}(\varphi_k)\}_k$. Notice that the relations
\eq{blopexp} and \eq{opkevsz} represent operator equations since Eq.
\eq{expval1} remains valid after the insertion of any other operator
in the path integral of the numerator at both sides.
 
It turns out to be useful to derive partial differential equations
for the $k$ dependence of the operator mixing matrix. Using
\begin{eqnarray}
   G_{\tilde n} (k-\delta k) &=&
   G_{\tilde n} (k) - {\delta k\over k} \beta_{\tilde n} (k,G) 
\end{eqnarray}
for an arbitrary infinitesimal change $\delta k$ of the scale $k$,
one can write
\begin{eqnarray}\label{betay}
S_{\tilde m\tilde n} (k-\delta k) &=&
       \frac{\delta G_{\tilde m} (k-\delta k)}{ 
             \delta G_{\tilde n} (\Lambda )}
=
    \sum_{\tilde\ell} \frac{\delta G_{\tilde m} (k-\delta k)}{
       \delta G_{\tilde\ell}(k)}
     \frac{ \delta  G_{\tilde\ell} (k) }{
            \delta G_{\tilde n} (\Lambda )} 
     =
   \sum_{\tilde\ell} \left\lbrack \delta_{\tilde m\tilde\ell}-
        \frac{\delta k}{k} \frac{ \delta \beta_{\tilde m}(k,G)}
{\delta G_{\tilde\ell} (k) }  \right\rbrack S_{\tilde\ell\tilde n} (k) .
\end{eqnarray}
Subtracting Eq. (\ref{mixmatdf}), dividing by $\delta k$ and taking
the limit $\delta k \to 0$, we find the  set of coupled differential
equations for the elements of the operator mixing matrix:
\begin{eqnarray}
\label{opmxset}
k\partial_kS_{\tilde m\tilde n}(k,\Lambda)
&=&\sum_{\tilde\ell}\frac{\delta\beta_{\tilde m}(k)}{\delta G_{\tilde\ell}(k)}
S_{\tilde\ell\tilde n}(k,\Lambda).
\end{eqnarray}
The scale dependence of the operator mixing matrix is governed by
the matrix 
\begin{eqnarray}\label{sensm}
k\Gamma_{\tilde n\tilde m}(k)&=&\frac{\delta\beta_{\tilde n}(k)}
{\delta G_{\tilde m}(k)}
\end{eqnarray}
determining the change of the slope of the RG trajectory due to the
variation of the actual
point of the parameter space it passes through. The matrix 
$\Gamma_{\tilde n\tilde m}(k)$ is a local feature of the RG trajectory
on the contrary to the sensitivity matrix.

\section{Parallel transport}\label{Parh}

\setcounter{equation}{0}

Now we show  that the operator mixing due to blocking
can be interpreted as the parallel transport of operators along the
RG trajectory with the connection $\Gamma_{\tilde n\tilde m}(k)$.

It is the main idea behind operator renormalization that operators are
searched for that reproduce the same vacuum expectation value at
different scales,
\bea\label{lepes}
{\int{\cal D}\phi_{k'}{\cal D}\chi_{k'}
 O_{\tilde n}(\varphi_{k'}+\chi_{k'})
e^{- S_{k'}\lbrack\varphi_{k'}+\chi_{k'}\rbrack}\over 
\int{\cal D}\phi_{k'}{\cal D}\chi_{k'}
e^{- S_{k'}\lbrack\varphi_{k'}+\chi_{k'}\rbrack}}
&=&{\int{\cal D}\phi_k{\cal D}\chi_k O_{\tilde n}(\varphi_k+\chi_k)
e^{- S_k\lbrack\varphi_k+\chi_k\rbrack}\over 
\int{\cal D}\phi_k{\cal D}\chi_ke^{- S_k\lbrack\varphi_k+\chi_k\rbrack}}.
\eea
This relation follows
from observing that the right hand side of Eq. (\ref{expval1}),
the expectation value of a given bare operator of the complete
cut-off theory, 
is independent of the choice of $k$, the way the modes are split
into the UV and IR classes. Eq. (\ref{lepes}) allows us to
identify the effect of the changing of the cut-off in the renormalized
operator,
\be\label{State2}
\{O_{\tilde n}(\varphi_{k'})\}_{k'}
={\int{\cal D}\xi\{O_{\tilde n}(\varphi_{k'}+\xi)\}_k
e^{- S_k\lbrack\varphi_{k'}+\xi\rbrack}\over 
\int{\cal D}\xi e^{- S_k\lbrack\varphi_{k'}+\xi\rbrack}},
\ee
for $k'<k$. The integration ${\cal D}\xi$ extends over the modes with
$k'<|p|<k$. The operator $\{O_{\tilde n}(\varphi_{k'})\}_{k'}$
is a linear superposition of the base operators $O_{\tilde n}(\varphi_{k'})$
as indicated in Eq. (\ref{blopexp}).

This result suggests a differential geometric interpretation of the
operator mixing, its identification with a certain parallel transport.
In particular, a $k$ dependent operator $O_k$ will be said
to be parallel transported in the scale $k$, 
\be
O_{k'}=P_{k\to k'}O_k
\ee
if its insertion into any expectation value yields a $k$-independent
result,
$\partial_k \langle O_k \rangle =0$. 
The mapping $P_{k\to k'}$ of the operators is obviously linear. 

Since the operator mixing is linear according to Eq. (\ref{blopexp}),
the parallel transport of an operator in the scale, i.e. the  parallel
transported operator $O_k$ can be characterized by introducing the covariant 
derivative of the operators,
\be\label{covder}
D_kO_k=(\partial_k-\Gamma)O_k
\ee
in such a manner that $D_kO_k=0$ for the parallel transport. 
Eq. (\ref{opmxset}) suggests the identification of 
the connection with the matrix $\Gamma$ of (\ref{sensm}).

The formal definition of the covariant derivative is the following.
The scale dependence in the expectation value $\langle O_k\rangle$
comes from two different sources: from the explicit $k$ dependence 
of the operator and the implicit $k$ dependence generated by the path
integration, by taking the expectation value.
The operator mixing ballances them. 
We introduce the covariant derivative by the relation
\be\label{covcon}
\partial_k\langle O_k\rangle=\langle D_kO_k\rangle,
\ee
requiring that the operator mixing generated by the connection
amounts to the implicit $k$  dependence of the expectation value.
Once the covariant derivative is known the parallel transport can be
reconstructed as
\be
P_{k\to k'}={\cal P}e^{\int_k^{k'}dk''\Gamma (k'')}
\ee
where $\cal P$ stands for the ordering according to the parameter
$k''$.

It is obvious that the connection $\Gamma$ is vanishing in the basis
$\{O_{\tilde n}\}_k$,
\bea
\partial_k\langle\sum_{\tilde n}c_{\tilde n}(k)\{ O_{\tilde n} \}_k\rangle
&=&\sum_{\tilde n}\partial_k c_{\tilde n}(k)\langle\{ O_{\tilde n}
\}_k\rangle
=\langle\partial_k\sum_{\tilde n}c_{\tilde n}(k)\{ O_{\tilde n} \}_k\rangle.
\eea
The connection can in principle be found in any other basis by
simple computation. As far as the basis $O_{\tilde n}(\varphi_k)$
is concerned it is simpler to check directly that (\ref{covder})
satisfies (\ref{covcon}). For this end we insert the arbitrary operator
\be
O_k=\sum_{\tilde m}c_{\tilde m}(k)O_{\tilde m}(\varphi_k)
={\underline c}(k){\underline O}
\ee
into (\ref{covcon}),
\be
\partial_k\left({\underline c}(k)\langle{\underline O}\rangle\right)
=\langle D_k{\underline c}(k){\underline O}\rangle,
\ee
and write, with Eq. (\ref{covder}),
\be
\partial_k{\underline c}(k)
\langle{\underline O}\rangle+
{\underline c}(k)\partial_k\langle{\underline O}\rangle
=\partial_k{\underline c}(k)\langle{\underline O}\rangle-
\langle{\underline O}\rangle_k{\underline{\underline\Gamma}}{\underline c}(k).
\ee
In order to prove (\ref{covcon}), we need the relation
\be\label{need}
-\partial_k\langle{\underline O}\rangle=
\langle{\underline O}\rangle{\underline{\underline\Gamma}}.
\ee
Using $S_{k=0}(G_{\tilde n}(k))$ as the generator function
for $\langle O_{\tilde n}\rangle$, the l.h.s. of
Eq. (\ref{need}) can be written as
\bea
{1\over\delta k}\left[{\delta S_0\over\delta G_{\tilde m}(k-\delta k)}
-{\delta S_0\over\delta G_{\tilde m}(k)}\right]
&=&{1\over\delta k} \left[
{\delta S_0\over\delta G_{\tilde m}(k-\delta k)}
- \sum_{\tilde n}
{\delta S_0\over\delta G_{\tilde n}(k-\delta k)}
{\delta G_{\tilde n}(k-\delta k)\over\delta G_{\tilde m}(k)}\right]
\nonumber\\
&=&{1\over\delta k}\sum_{\tilde n}
{\delta S_0\over\delta G_{\tilde n}(k-\delta k)}
\left[\delta_{\tilde n\tilde m}
-{\delta G_{\tilde n}(k-\delta k)\over\delta G_{\tilde m}(k)}\right]
={1\over k}\sum_{\tilde n}
{\delta S_0\over\delta G_{\tilde n}(k)}
{\delta\beta_{\tilde n}(k)\over\delta G_{\tilde m}(k)},
\eea
in the limit $\delta k\to0$. According to (\ref{opmxset}), 
the last line agrees with the r.h.s. of Eq. (\ref{need}). 

Let us turn now to the usual quantum field theoric problem without
 inhomogeneous external sources, that have only been introduced as
 technical tools, i.e. take the limit $g_{\tilde n}(k)\to0$. Then
the operator mixing matrix and the connection reduce to
\be
S_{\tilde m\tilde n}(k,\Lambda)=\delta(x_m-x_n)s_{mn}(k,\Lambda) ,
     \qquad
\Gamma_{\tilde m\tilde n}(k)=\delta(x_m-x_n)\gamma_{mn}(k),
\ee
resp. with the coordinate independent matrices
\begin{eqnarray}
\label{Yijsij}
  s_{mn} (k,\Lambda) = \frac{ \partial g_m (k)}{ \partial g_n (\Lambda )},
       & \qquad &
 k \gamma_{mn} (k) = \frac{ \partial \beta_m (k)}{ \partial g_n (k) } 
\end{eqnarray}   
and the RG equations for the operator mixing matrix take the form
\begin{eqnarray}
 \partial_ks_{mn}(k,\Lambda)&=&\sum_\ell \gamma_{m\ell}(k)
s_{\ell n}(k,\Lambda).
\end{eqnarray}

\section{RG flow and universality}\label{Classific}

\setcounter{equation}{0}

It has been found that the operator mixing, the problem of
keeping the expectation values cut-off independent can be
handled by a linear transformation, the parallel
transport of the operators. We shall show in this section 
that the salient features of the RG flow can be recovered
from this parallel transport alone.

The scaling combinations of the coupling constants are introduced
traditionally in the vicinity of a fixed point $G^*_{\tilde m}$ in
the space of the coupling constants which have been made dimensionless
by the help of the cut-off $k$. The basic assumption of the RG strategy
is that the evolution equations can be linearized around the fixed points,
\be
\beta_{\tilde n}\approx\sum_{\tilde m}
\Gamma^*_{\tilde n\tilde m}(G_{\tilde m}-G^*_{\tilde m}).
\ee
The relevant, marginal and irrelevant coupling constants are the
superpositions
\be
G^{sc}_{\tilde n}=\sum_{\tilde m}{\bar c}_{\tilde n\tilde m}
\left(G_{\tilde m}  -  G^*_{\tilde m} \right)
\ee
made by the left eigenvectors of $\Gamma$,
\be
\sum_{\tilde m} {\bar c}_{\tilde n\tilde m}
\Gamma^*_{\tilde m\tilde r} =\alpha_{\tilde n}{\bar c}_{\tilde n\tilde r},
\ee
with $\alpha_{\tilde n}<0$, $\alpha_{\tilde n}=0$ and
$\alpha_{\tilde n}>0$, resp. The scale
dependence of the scaling coupling constants $G^{sc}_{\tilde n}$ is
\be\label{scl}
G^{sc}_{\tilde n} \sim  k^{\alpha_{\tilde n}}.
\ee

Let us consider a local operator of the form
\be
O(\phi(x)) = \sum_n b_n O_n ( \phi(x))
\ee
where $O_n(\phi(x))$ is the product of the terms
$\partial_{\mu_1}\cdots\partial_{\mu_\ell}\phi^m(x)$.
It will be necessary to separate for the possible scale dependent
rescaling factors. For this end we introduce the norm
\be
||O||=\sqrt{\sum_n b_n^2  }
\ee
and adopt the convention that the coupling constants
$G_{\tilde n}(\Lambda)$ of the bare action always multiply
local operators with unit norm, $||O_{\tilde n}(\Lambda)||=1$.
As long as we consider local operators the norm
defined above is sufficient and no index with continuous
range is needed in its definition. It will be useful to consider
the normalized operator flow,
\be
\overline{O}={O\over||O||},
\ee
in addition to the original one, $O$.

The scaling operators
\be
O^{sc}_{\tilde n}=\sum_{\tilde m}c_{\tilde n\tilde m}O_{\tilde m}
\ee
are obtained by means of the right eigenvectors of $\Gamma^*$,
\be
\sum_{\tilde m}\Gamma^*_{\tilde r\tilde m}c_{\tilde m\tilde n}
=\alpha_{\tilde r}c_{\tilde r\tilde n},
\ee
satisfying the conditions of completeness
$c\cdot{\bar c}=1$ and orthonormality ${\bar c}\cdot c=1$.
The coupling constants of the action
\be
S_k=\sum_{\tilde n}G_{\tilde n}(k)\overline{O^{sc}_{\tilde n}(\phi_k)}
\ee
obviously follow (\ref{scl}). The operator
\be
O=\sum_{\tilde n}b_{\tilde n}\overline{\{O^{sc}_{\tilde n}\}_k},
\ee
written at scale $k$ in this basis yields the parallel transport trajectory
\be
\{O\}_{k'}=\sum_{\tilde n}b_{\tilde n}
\{\overline{\{O^{sc}_{\tilde n}\}_k}\}_{k'}
=\sum_{\tilde n}b_{\tilde n}
\left({k'\over k}\right)^{\alpha_{\tilde n}}
\overline{\{O^{sc}_{\tilde n}\}_k}
\ee
in the vicinity of the fixed point.

The question we explore now is what information does the flow $\{O\}_k$
contain about the importance
of certain interactions as the function of the observational scale $k$.
Let us start with the remark that according to \eq{blopenot}
the fixed point of the blocking relation, where the action, expressed
in terms of the dimensionless coupling constants, is scale independent
agrees with the fixed point of the operator blocking.
The linearization around the fixed points renders the critical
exponents of the coupling constants and the scaling operators equivalent
since the left and the right spectrum of $\Gamma^*$ agree.
The operators whose structure converges and changes by an overall
factor only are the scaling operators. The corresponding critical
exponents can be read off from the evolution of their norm.

The concept of universality stands for
the independence of the dimensionless quantities of the long distance
physics from the initial value of the irrelevant coupling constants in
the UV scaling regime.
The operator $\{O_{\tilde n}(\varphi_k) \}_k$ constructed from the
local terms of the bare action by parallel transport represents
the influence of the bare coupling constants on the physics of the
scale $k$. Since the action is dimensionless
only relevant operators have non-vanishing parallel transport flow
at finite scale according to Eq. \eq{blopenot}, the parallel transport
of the irrelevant operators which are made dimensionless by the
running scale $k$ vanishes. Naturally this does not mean
that the effective theories are renormalizable since a renormalizable
(relevant) operator parallel transported down from the cut-off mixes with
non-renormalizable (irrelevant) ones, as well.

The operator flow is particularly well suited for the studies of models
whose RG flow visits different scaling regimes. In general, any 
renormalizable theory without manifest scale invariance possesses at 
least two scaling regimes, one around the UV and another one at the 
IR fixed points which are separated by a crossover at the intrinsic scale 
of the theory, $k=k_{cr}$. For models with mass gap the IR scaling regime 
is trivial, i.e. the relevant operators are Gaussian. Imagine a model with 
dynamically generated intrinsic scale, e.g. with spontaneous symmetry 
breaking or condensation or dimensional transmutation, where the IR instability
generates non-trivial scaling laws and non-Gaussian relevant coupling 
constants appear in the IR scaling. Let us suppose
that there is a non-renormalizable operator $O$ which
becomes relevant in the IR regime and consider its
parallel transport globally, from the UV to the IR fixed point 
\cite{Alex}. Since $O$ is non-renormalizable $||\{O\}_k||$
decreases as $k$ is lowered in the UV reflecting the diminishing
importance of a non-renormalizable operator well below the
cut-off. But after having crossed the crossover the flow reflects the 
properties of a relevant operator, i.e. $||\{O\}_k||$ increases as $k$ 
is further lowered in the IR  scaling regime. Now it becomes a 
competition between the UV suppression and the IR enhancement
to form the final sensitivity on the cut-off scale parameter.
Since the IR increase of the norm is usually fed by IR or
collinear divergences the length of the scaling regimes are
determined by $\Lambda/k_{cr}$ and $Lk_{cr}$ where $L$ is the
size of the system, an IR cut-off. If the coherence is not
lost for sufficiently large distances then the condition
$||\{O\}_k||=1$ can be reached at low enough $k$ for any
value of the UV cut-off, i.e. the IR instability can 
 make the otherwise weak sensitivity on the short distance
physics strong.

Finally, few words are in order about the possibility of
determining the sensitivity matrix. There is a {\em direct} method 
by solving the set (\ref{opmxset}).
 Namely, the right hand sides of Eq. 
(\ref{couWHeq}) are analytic expressions for the beta-functions that 
can be differentiated analytically, and the numerical values of the 
connection matrix then easily computed and used as input for Eq. 
(\ref{opmxset}). There is also an {\em indirect} method according to the
definition (\ref{mixmatdf}). One has to solve the Wegner-Houghton
equations (\ref{couWHeq}) for the blocked sources as functionals of their
initial values at the scale $\Lambda$, and differentiate the solution
w.r.t. these initial conditions. This latter method was used in
Ref. \cite{Alex} to show that the $\phi^4$ model in the
phase with spontaneously broken symmetry does in fact possess
a non-renormalizable relevant operator in the IR scaling regime.

\section{RG and Renormalized Perturbation Expansion}
\label{Compar}
\setcounter{equation}{0}
The comparison of the renormalisation of the composite
operators presented above with the usual operator mixing
obtained in the framework of the renormalized perturbation expansion
serves two goals. First, it shows that the usual operator mixing
represents the evolution of the composite operators towards
the UV direction. Second, it helps to understand an
apparent paradox, namely that non-renormalizable operators
can be "renormalized" within a renormalizable theory. The
composite operator renormalization stands for the program of finding
the counterterms which renders the Green's functions even with the given
composite operator insertions finite as the cut-off is removed.
Since the finiteness of the Green's functions implies the finiteness of the
partition functions where the composite operators are introduced
with a source term in the action, as in Eq. \eq{blact}, the completion
of this program would amount to the renormalization of theories where
the composite operators appear in the action.

\subsection{Renormalized Perturbation Expansion}
Let us start with the bare action (\ref{bareact}),
\begin{eqnarray}
  S \lbrack \phi, G_{\tilde n} \rbrack 
            &=&
     \sum_{\tilde n} G_{\tilde n}  O_{\tilde n} ( \phi ) .  
\end{eqnarray}
For the sake of simplicity, we neglected the subscript $\Lambda$  here,
but the dependence on the external sources $G_{\tilde n}$ is made explicit;
 $O_{\tilde n} (\phi )$ is  a complete set of normalized
bare operators. The UV and IR momentum
cut-offs, $\Lambda $ and $k$ are assumed to be introduced for the
field variable $\phi (x)$ in order to achieve a better comparison
with the RG method. Let
us separate the zero modes of the sources, i.e. the bare coupling
constants $g_n$, $G_{\tilde n}=
g_n + g_{\tilde n}$. Furthermore, specify $O_{(1|1)} (\phi (x) )= -\frac{1}{2}
\phi (x) \Box \phi (x)$,  $O_1 (\phi (x) )=\phi (x)$,
$g_{(1|1)} = Z_\phi$, $g_1=0$. Then,  $G_{\tilde 1} = g_{\tilde 1}  =
-j(x)$ is the external current coupled to the bare field.

The corresponding quantum field theory is defined 
 by the generating functional 
\begin{eqnarray}
\label{Zbare}
      Z\lbrack  G_{\tilde n}   \rbrack 
     &=&
        \frac{
  \int {\cal D} \phi e^{ - S \lbrack \phi, G_{\tilde n}\rbrack }
              }{
    \int {\cal D} \phi e^{ - S \lbrack \phi;
  g_n \rbrack}
                } .
\end{eqnarray}
Then, we obtain the  generating functionals $W \lbrack
 G_{\tilde n} \rbrack
=\ln Z \lbrack G_{\tilde n} \rbrack$
and $\Gamma \lbrack \varphi,  $ $ G_{{\tilde n}, n \neq 1} \rbrack =
 - W \lbrack  G_{\tilde n}  \rbrack
+ \int d x j(x) \varphi(x)$ of the  connected
 and 1PI Green's functions,
resp., with $\varphi (x) = \delta W/\delta j(x)$.

The  insertion of  the operator $O_{\tilde n} (\phi )$  in the 1PI Green's
functions is obtained via functional derivation w.r.t. the
corresponding source $G_{\tilde n}$, and the identities
\begin{eqnarray}
  \left\langle O_{\tilde n} ( \phi  ) \right\rangle_{ 1PI}
          &=&
        \left.
 - \frac{1}{Z} \frac{ \delta Z}{\delta G_{\tilde n}} 
      \right|_{0}
              =
   \left.
     - \frac{ \delta W}{ \delta G_{\tilde n} }
       \right|_{ 0}
           =    
   \left.
 \frac{ \delta \Gamma }{ \delta G_{\tilde n} }
   \right|_{ 0}
\end{eqnarray}
hold for $n \neq 1$. The subscript $\ldots |_0$ stands for $g_{\tilde
  n} \equiv 0$ (and $\varphi =0$ for $\Gamma$).
 Owing to its definition via the bare action, this vacuum
expectation value is expressed in terms of the bare couplings.

The theory defined above is non-renormalizable. 
One has to satisfy infinitely many
renormalization conditions in order to fix the renormalized values
$g_{nR}$ of the
infinitely many coupling constants. This is equivalent to specifying
a particular RG trajectory in the RG approach.
The theory with non-renormalizable coupling constants
does not allow the removal of the cut-off, i.e. the
extrapolation towards short distances is problematic.
But as long as the properties far away from the cut-off
towards the IR direction are concerned and the
possible non-trivial effects of the IR scaling regime
\cite{Alex} are neglected, the values of the
non-renormalizable coupling constants at the cut-off
effect the overall scale of the theory only.
When considering dimensionless quantities this scale
drops out and the non-renormalizable coupling constants
can be set at the cut-off in an arbitrary manner. As a result,
there is no problem to construct the operator mixing below
a sufficiently high  cut-off. One should bear in mind that even
the renormalizable theories contain non-renormalizable operators,
the regulator. In fact, the comparison of a theory with different
regularizations shows that the regulators
amount to a set of irrelevant operators when written in the action.
These regulator terms have tree-level fine-tuning which,
according to the universality, is sufficient to keep the cut-off
independent dynamics fixed. In what follows we take the usual
point of view and the regulators will not be represented in the
action.

The renormalization conditions enable one to
rewrite the  bare action as the sum of the renormalized terms and
that of infinitely many counterterms
\begin{eqnarray}
\label{Sbaren}
      S \lbrack \phi, G_{{\tilde n} } \rbrack 
   & = &  
      \sum_{\tilde n} \left(  G_{{\tilde n}R}
       +    c_{\tilde n} \lbrack G_{{\tilde m} R}
   \rbrack    \right)
  O_{\tilde n} (\phi )     
      \equiv
S_R \lbrack  \phi , G_{{\tilde n}R} \rbrack
\end{eqnarray} 
in the framework of the RPE.
Here we introduced the renormalized  sources via  $ G_{\tilde n} =
G_{{\tilde n}R}  + c_{\tilde n} \lbrack G_{{\tilde m}R} \rbrack$.
As mentioned above, all but finite counterterms influence an
overall scale factor only of the theory according to the universality.
The cut-off will be kept arbitrary large but finite and fixed.
Below the notation $c_{(1|1)}\lbrack G_{{\tilde m}R} \rbrack = Z_\phi
-1$ and the renormalization condition $G_{(1|1)R}  \equiv 1$
are used. The coefficients of the counterterms $c_{\tilde n}$ are
functionals of the renormalized sources $G_{{\tilde m}R} $.
Without loss of generality we can assume that
all  external sources are cut off at some momentum $\Lambda_s \ll
\Lambda$. Then, the counterterms remain local similarly to the case
with constant external sources  \cite{Col84} and
the coefficients  $c_{\tilde n}$ are only affected
by the zero modes of the sources, i.e. they are independent of the
spacetime coordinate $x$ and are functions of the coupling constants
$g_{nR}$ and the  UV and IR cut-offs, $\Lambda$ and $k$, resp.
Then, the relations $g_{nR} + c_{n}  ( g_{mR})  =  g_{n}$ and
$  g_{{\tilde n}R}  \equiv   g_{\tilde n} $ hold.
For $n=1 $ these yield $-g_{\tilde 1}  \equiv j(x) =  j_R (x)$ and
$   g_{1} = c_1$ for the renormalization condition $g_{1R}=0$.
We see that $g_{\tilde n} \equiv 0$ implies $g_{{\tilde n}R} \equiv 0$
and vice versa.

The renormalized generating functional $Z_R$ considered as the functional of the
renormalized sources,
\begin{eqnarray}
 Z_R \lbrack  G_{{\tilde n} R}  \rbrack
     &=&
  \frac{
 \int {\cal D} \phi (x)
 e^{- S_R \lbrack \phi , G_{{\tilde n} R}\rbrack   }
        }{
  \int {\cal D} \phi (x)
e^{- S_R \lbrack \phi ;  g_{n R} \rbrack }
 }    
       =
 Z \lbrack G_{\tilde n} \rbrack 
\end{eqnarray}
is the generating functional of the Green's functions with renormalized
composite operator insertions. The following equations hold:
\begin{eqnarray}
   W_R \lbrack  G_{{\tilde n} R} \rbrack
      = \ln  Z_R \lbrack  G_{{\tilde n} R}  \rbrack
         &=&
          W \lbrack  G_{\tilde n}  \rbrack ,
\end{eqnarray}
and 
\begin{eqnarray}
  \Gamma_R  \lbrack  \varphi, G_{{\tilde n} \neq {\tilde 1},R}  \rbrack
  &=&
 - W_R \lbrack G_{{\tilde n} R}  \rbrack 
   + \int dx j_R(x)\varphi(x)
         =
     - W \lbrack G_{\tilde n}  \rbrack
 + \int dx j(x) \varphi(x)    
          =
       \Gamma  \lbrack  \varphi, G_{{\tilde n} \neq {\tilde 1}} \rbrack ,
\end{eqnarray}
where $ \varphi(x) = \delta W_R /\delta j_R (x) = \delta W / \delta j (x)$.

The renormalized composite operator insertion 
$\left\lbrack O_{\tilde n} ( \phi ) \right\rbrack_R$ $(n \neq 1)$ is obtained
 by functional derivation
w.r.t. the renormalized external source $G_{{\tilde n}R } $:
\begin{eqnarray}
  \left\langle \left\lbrack O_{\tilde n} ( \phi  ) \right\rbrack_R
      \right\rangle_{1PI}
    &=&
   - \left. \frac{1}{Z_R} \frac{
 \delta Z_R \lbrack  G_{{\tilde m} R} \rbrack }{
      \delta G_{{\tilde n}R} } 
      \right|_{ 0}
          =   
   - \left. \frac{
 \delta W_R \lbrack  G_{{\tilde m} R} \rbrack }{
      \delta G_{{\tilde n}R}  } 
      \right|_{ 0}
       =
 \left.
   \frac{\delta \Gamma_R \lbrack \varphi , G_{{\tilde m}\neq {\tilde
         1},R}
 \rbrack  }{
      \delta G_{{\tilde n}R}  }
  \right|_{ 0}
       .
\end{eqnarray}
Now we find the following relations
\begin{eqnarray}
   \left.
  \frac{ \delta \Gamma_R \lbrack \varphi , G_{{\tilde m}\neq {\tilde
        1},R} 
\rbrack
        }{
   \delta G_{{\tilde n}R}  
        }
    \right|_{ 0}
  & = &
      \left.
   \frac{ \delta \Gamma \lbrack \varphi , G_{{\tilde m} \neq {\tilde
         1}}
 \rbrack
        }{
   \delta G_{{\tilde n}R}  
        }
      \right|_{ 0}
         = 
  \sum_{{\tilde r}\neq {\tilde 1}} 
        \left.  \frac{\delta   G_{\tilde r}
                       }{
         \delta G_{{\tilde n}R}   }
        \right|_{0} 
        \left.
   \frac{ \delta \Gamma \lbrack \varphi, G_{{\tilde m} \neq {\tilde
         1}}
   \rbrack  }{
            \delta  G_{\tilde r}   }
 \right|_{0} .
\end{eqnarray}
Making use of the derivative
\begin{eqnarray}
     \left.  \frac{ \delta  G_{\tilde m} }{
         \delta G_{{\tilde n}R}   }
        \right|_{0} 
     & =&
   (Z^{-1} )_{mn} \delta ( x-y) 
\end{eqnarray}
with the operator mixing matrix
\begin{eqnarray}
\label{defzmn}
  ( Z^{-1})_{mn}  &=&
    \delta_{mn} + \frac{ \partial c_m (g_{rR} )}{ \partial g_{nR} }
\end{eqnarray}
(for $n, m \neq 1$), one finds 
\begin{eqnarray}
\label{renoper2}
  \left\lbrack O_{\tilde n} ( \phi  ) \right\rbrack_R 
      &=&
    \sum_{ m\neq  1} 
           O_{\tilde m} (\phi  )  
\left(  Z^{-1} \right)_{mn}  
\end{eqnarray}
for $n\neq 1$. This can be extended to $n,m=1$ by defining 
$ ( Z^{-1})_{1m} =\delta_{m1}$, $ ( Z^{-1})_{n1} =\delta_{1n}$.
It is worthwhile mentioning that the operator mixing matrix turned out
local in  spacetime coordinates, i.e. it is independent of the
momentum at which the composite operator insertion is taken. The
matrix $Z_{nm}$ is just the transposed of that used in Ref. \cite{Col84}.

\subsection{Comparison of the two schemes}\label{comRGper}
We compare now the notion of renormalized operator
in the perturbative approach with the notion of blocked operator
in the RG framework. Table \ref{parallel} summarizes the
formal similarities between the two approaches.

The RG approach keeps the UV cut-off $\Lambda$ fixed
and uses a decreasing IR cut-off $k$, so that the RG trajectories are
passed  towards the IR limit $k \to 0$. On the contrary, the
couplings are defined at some low-energy scale $k=\mu \ll \Lambda$
and the UV cut-off is shifted towards infinity in the
RPE, and the RG trajectories are
followed just in the opposite direction,
towards large momenta. The RG approach
reproduces the perturbative results for the ordinary Green's functions
in the UV scaling regime \cite{Col84,Polch84} up to powers and
logarithms of $\mu/\Lambda$ which are vanishing in the asymptotic
limit $\Lambda\to\infty$.

It is easy to see that
\be
\sum_{\tilde n}G_{\tilde n}(\Lambda)\{O_{\tilde n}\}_k=
\sum_{\tilde n}G_{\tilde n}(\Lambda)
{\delta S_k[\phi_k]\over\delta G_{\tilde n}(k)}
\ee
gives the blocked action with cut-off $k$ in the order
${\cal O}(G(\Lambda))$, the blocking of the action and the
operators agree in the linearized level, i.e. they share the
fixed points and the critical exponents.
The Legendre transformed effective action with the IR cut-off
$k \neq 0$ \cite{Exact,Polon} describes the effective theory after the
high-frequency modes have been eliminated, similarly to the
blocked action for the low-frequency modes.

The RPE deals with the
effective action in the limit $k \to 0$, $\Lambda \to \infty$.
Since the bare couplings $g_n $ and the renormalized couplings
$g_{nR}$ are the analogues of $g_n (\Lambda )$ at the UV cut-off
scale and of the blocked couplings $g_n (k)$,  resp., the
operator mixing matrix $s_{nm}
= \partial g_n (k) / \partial g_m (\Lambda )$ defined in the RG
approach is just the analogue of the  matrix $Z_{nm}$ defined via
$(Z^{-1})_{mn} = \partial g_m/\partial g_{nR}$ in the RPE.
This analogy will be demonstrated in Sect. \ref{Opmixing}
on the equivalence of the
one-loop perturbative results with those obtained by the RG method
in IMA for a few elements of the operator mixing matrix in the
particular case of $\phi^3$ theory in dimension $d=6$.

The analogues of the blocked operators
$\{ O_{\tilde n} (\varphi_k ) \}_k $ are  not used in the RPE. In the
latter one seeks the operator  at the scale of the
UV cut-off that reproduces the expectation value of an operator
given at the renormalization scale. The renormalized operator
$\lbrack O_{\tilde n} (\phi  )\rbrack_R$ satisfying this requirement is
the analogue of the operator $\lbrack O_{\tilde n} \rbrack_k (\phi )$
introduced in the RG approach by means of inverse blocking. 

It is the basic advantage of the RG approach that a non-perturbative answer
can be obtained with its help on operator mixing, whereas also the
UV finite pieces of the operator mixing matrix are automatically determined.
Furthermore, the RG approach enables one to get  a deeper insight
in the reason of operator mixing as a natural consequence of
reproducing the same vacuum expectation values in the bare theory and
in the blocked one, or in other words as a direct consequence of
integrating out degrees of freedom.

\section{Operator mixing in $\phi^3_6$ theory}\label{Opmixing}
\setcounter{equation}{0}
The relation between the composite operator renormalization in
the RG scheme and the operator mixing of the RPE
is demonstrated in this section in the case of a simple scalar model.

\subsection{RG method}\label{OpmRG}
We start with the determination of the blocked operators in
the framework of the $\phi^3$ theory in dimension $d=6$ in the
independent mode approximation (IMA) of the next-to-leading order of
the derivative expansion and show that both the direct and
the indirect methods lead to the same results.

Let us include as base operators $O_{\tilde n} (\varphi_k  )$
the derivative operators
\begin{eqnarray}
\label{derops}
  D_{(0|1) } (\varphi_k  ) = - \Box \varphi_k , 
         \quad
  D_{(1|1)} (\varphi_k ) =  - \frac{1}{2} \varphi_k \Box \varphi_k ,
            \quad
  D_{(0|2)} (\varphi_k )    
   =  - \frac{1}{2} \Box \varphi_k^2 
\end{eqnarray}
and the local potential
\be
V(\varphi_k)=\sum_{\ell=0}^\infty g_\ell(k){\varphi_k^\ell\over\ell!}
\ee
into the Ansatz for the blocked action:
\begin{eqnarray}
\label{genaction}
 S_k &=&
       \int d^d x \left\lbrack
    Z_{(0|1)} (k) D_{(0|1) } (\varphi_k )
   + Z_{(1|1)} (k) D_{(1|1)} (\varphi_k )
  +Z_{( 0|2)} (k)   D_{( 0|2)} (\varphi_k )
  + V (\varphi_k ) 
         \right\rbrack.
\end{eqnarray}
The bare $\phi^3$ theory is specified by
the bare couplings $g_2(\Lambda )= m^2$, $g_3 (\Lambda)=\lambda$,
$g_{\ell \ge 4} (\Lambda ) =0$, $Z_{(0|1)}(\Lambda )=0$,
$Z_{(1|1)}(\Lambda )=1$,  and  $Z_{(0|2)}(\Lambda ) = -\hf$.

The choice of operators in (\ref{genaction}) means that the field
dependence of the wave
function renormalization is not taken into account.
Terms with higher order derivatives of the field  are also neglected, but the
 extension of the operator basis
 is straightforward.
Other
operators of the  discussed types can be expressed as linear
 combinations of those included in the basis,
e.g.
\begin{eqnarray}
\label{dfidfi}
  - \partial_\mu \varphi_k \cdot \partial_\mu \varphi_k &=& 
D_{( 0|2)} (\varphi_k )
   - 2 D_{( 1|1)} (\varphi_k ) .
\end{eqnarray}

Having derived the explicit forms of the right hand sides of
Eqs. (\ref{WHeq}) and (\ref{opmxset}), the limit $g_{\tilde n} (k) \to 0$
can already be taken  at the beginning of the calculation except for
the couplings multiplying the operators $\Box \varphi_k (x)$
and  $\Box \varphi_k^2 (x)$. The corresponding terms in the action
 would yield
pure surface terms and therefore, their effects on operator mixing can
only be kept track if the corresponding inhomogeneous sources are
replaced
 by constants only at the end of the calculation. 

A further remark is in order here. Namely, we have formulated the whole
procedure in the Wegner-Houghton framework with a sharp cut-off and
used derivative expansion. It
is, however, well-known that the sharp cut-off introduces undesirable
singularities due to the derivatives of the step like cut-off
function \cite{Morris}. This may also cause our method in its present
form with sharp cut-off to fail in correctly describing the
renormalization of the derivative operators.
We do not see, however, any objections
to  reformulate our method of treating composite operator
renormalization  using a smooth cut-off on the base of Polchinski's
equation \cite{Polch84}.

\subsubsection{Indirect method}\label{indiropmx}

For the indirect determination of the operator mixing matrix
the following steps should be performed:
\begin{enumerate}
\item {\bf Determination of the blocked couplings.}
Our method to establish the coupled set  of differential equations
(\ref{couWHeq}) 
for the couplings from Eq. (\ref{WHeq}) is
similar to that used in \cite{Fra85}. Namely, we substitute
$\varphi_k (x) = \varphi_0 + \eta (x)$ where $\varphi_0$ is an
arbitrary constant, 
expand both sides of
Eq. (\ref{WHeq}) in Taylor series w.r.t. the inhomogeneous piece $\eta
(x)$, and compare the coefficients of the corresponding operators $
O_{\tilde n} (\eta  )$. Since only operators with second derivatives are
considered, it is sufficient to terminate the expansion at the
quadratic terms. 
In Appendix \ref{RGcoup}, the set of coupled first order differential
 equations  
 (\ref{potbl}), (\ref{sig1a1}),
(\ref{sig1a2}), and (\ref{sig2c1}) is obtained
 for the
blocked couplings. These equations are then solved in independent mode
approximation analytically  in Appendix \ref{coupIMA}.

\item {\bf Determination of the operator mixing matrix in IMA} through
differentiating the solutions for the blocked couplings w.r.t. the
initial values of the various  couplings at the scale $\Lambda$. This
step is discussed in detail in App. \ref{coupIMA}. The orders of
magnitude of the elements of the operator mixing matrix
w.r.t. the UV cut-off $\Lambda$ are indicated in Table \ref{YijIMA}.

\end{enumerate}

\subsubsection{Direct method}\label{diropmx}

As byproducts, the right hand sides of  Eqs. (\ref{potbl}), (\ref{sig1a1}),
(\ref{sig1a2}), and (\ref{sig2c1}) provide us the exact  analytic
expressions for the beta-functions, $\beta_n (k)$ (see the beginning
of Appendix  \ref{senmat}).
 Thus, analytic
expressions can be obtained for the elements of the
connection $\gamma_{nm}$ by differentiating the appropriate
beta-functions w.r.t. the appropriate coupling constants without any
 approximation. The results
are summarized in Appendix \ref{senmat} and the non-trivial matrix
elements indicated in Table \ref{sensim}.
This matrix is the input for Eqs. (\ref{opmxset}) to  the direct determination
of the operator mixing matrix.

Eqs. (\ref{opmxset}) can be rewritten as integral equations
 in a rather compact form
introducing the coloumn vectors ${\underline s}_n$ with the elements
$s_{mn}$  (with the row index $m$) and the connection matrix ${\underline
  {\underline \gamma}}$  (with the elements $\gamma_{lm}$):
\begin{eqnarray}
\label{opmxvc}
   {\underline s}_n (k) &=& {\underline s}_n (\Lambda )
       - \int_k^\Lambda 
       d\kappa {\underline {\underline \gamma}} (\kappa ) 
   {\underline s}_n (\kappa ) .
\end{eqnarray}
This means that all the coloumn vectors ${\underline s}_n (k)$ satisfy
the same ordinary first order linear differential  equation, only the
initial conditions $ {\underline s}_n (\Lambda )$ are different for
them. The analytic expressions for the elements of the connection
matrix  (see Appendix \ref{senmat}) should be used as input.

Here we determine the operator mixing matrix analytically in IMA.
For this approximation, all the coupling
constants in the explicit expressions for the elements of ${\underline
{\underline \gamma}}$ and the coloumn vector ${\underline s}_n
(\kappa)$ should be replaced by their bare values
on the r.h.s. of Eq. (\ref{opmxvc}). For the $\lambda \phi^3$ theory one has
$Z_{(1|1)} (\Lambda ) =1$,  $Z_{(0|1)} (\Lambda)=0$,
$Z_{(0|2)}(\Lambda )= -\hf$, $g_2 (\Lambda )=m^2$,
$g_3 (\Lambda )=\lambda$, $g_{\ell \ge 4} (\Lambda ) =0$. Since there is
no operator mixing at the scale $\Lambda$, the initial conditions are
$s_{mn} (\Lambda ) =\delta_{mn}$. Then,
Eqs. (\ref{opmxvc}) take the forms
\begin{eqnarray}
\label{opmxvcIMA}
   s^{IMA}_{mn} (k) &=& \delta_{mn}
       - \int_k^\Lambda 
       d\kappa  \gamma^{IMA}_{mn} (\kappa ) , 
\end{eqnarray} 
and their
solutions can be expressed in terms of
the
 integrals in 
App. \ref{integrals}. Using the results of App. \ref{senmat},
it has  been checked that the solutions are just the operator mixing
coefficients  found in App. \ref{coupIMA} 
by the indirect method previously.

\subsection{Perturbative approach}\label{Opmper}

In App. \ref{per1loop} we determine perturbatively
  the operator mixing coefficients $(Z^{-1})_{nm}$
  for $\phi^3$ theory in dimension $d=6$  in one-loop
approximation.  The inverse of  $(Z^{-1})_{nm}$
can directly be compared with the results obtained for the matrix $s_{mn}$
by means of  the RG approach in IMA in Sect. \ref{OpmRG}. 
The inversion of the
matrix results in the change of the sign of the terms of ${\it o}
(\hbar )$ of the matrix elements. Expressing the perturbative results in
terms of the loop integrals given in App. \ref{integrals}, it is easy
to recognize that  $Z_{nm} = s_{nm} (k=0)$ in the above mentioned
 approximations. This agreement illustrates how the perturbative
 approach is related to the RG approach, that was discussed in Sect.
\ref{comRGper}.

\section{Summary}\label{summary}
In the toy model of a zero-dimensional field theory
the operator mixing has been explained as the natural consequence of
integrating out degrees of freedom.
Then, the notion of blocked operators is defined in one-component
scalar field theory through the requirement of reproducing the same
vacuum expectation value in the bare theory and in the blocked
(effective) one. The blocking procedure proposed by
Wegner and Houghton has been used, i.e. the high-frequency degrees of
freedom  were integrated out in infinitesimal momentum shells sequentially
using a sharp moving cut-off.

It is shown that the blocking of operators introduced in this paper
satisfies the (semi)group property. Differential equations are derived
for the elements of the operator mixing matrix that should be solved
simultaneously with the Wegner-Houghton equations.
It is found that the blocking of operators corresponds to
parallel transporting them along the RG trajectory, a flat,
one-dimensional manifold.
The scale dependence of the operator mixing matrix is governed by the
connection, showing the changes of the beta-functions due to
infinitesimal changes of  the couplings at a certain scale $k$.
It is also shown that
the eigenoperators of the connection are the local
scaling operators in any scaling regimes. Thus, solving the eigenvalue
problem of the connection enables one in principle to detect different
scaling regimes and find the corresponding relevant operators.

The limitations are, of course, the validity region of the
Wegner-Houghton equation due to occurring a non-trivial saddle point,
and the usage of the sharp cut-off together with the gradient
expansion.
As to the latter technical problem, a generalization of operator
blocking to a smooth cut-off approach seems to be possible by
including the appropriate cut-off terms into the action. The
reformulation of the whole issue in the framework of
renormalization in the internal space \cite{Polon}
would solve both of the above mentioned problems and
do also for generalization to gauge theories.

The differential RG approach and the perturbative approach for
operator renormalization are compared in detail. It is explained that
the renormalized operator (used in perturbative terms)
corresponds to choosing an operator at the renormalization scale
and looking for an operator at the UV scale (tending to infinity)
that reproduces the vacuum expectation value of the chosen operator.
On the particular example of $\phi^3$ theory in dimension $d=6$
the agreement of the results of the RG approach in IMA with  
the one-loop perturbative ones has been illustrated for the elements
of the operator mixing matrix in a truncated basis of operators.

The main advantage of the flow equations for the sensitivity matrix presented
is that by their help the renormalization of composite operators
can be performed beyond the perturbative regime. The method has
the power to go beyond the Independent Mode Approximation by solving
the flow equations numerically. A comparison of the non-perturbative results
with the perturbative ones e.g. for the well-known $\phi^3$ theory would
certainly be interesting. Various models with non-trivial IR scaling could be
analysed down to the scale where the IR instabilities responsible for the
non-trivial scaling occur.

\acknowledgements One of the authors (K.S.) expresses his gratitude
to the Humboldt Foundation for the follow-up
grant, to  W. Greiner and G. Soff for their kind hospitality, and thanks
I. Lovas, Z. Nagy, G. Plunien, and R. Sch\"utzhold for the useful discussions.
Valuable discussions with N. Sadooghi are appreciated, too.
This work has been supported by the grants OTKA T023844/1997, T29927/98, NATO
SA(PST.CLG 975722)5066, DFG-MTA 436UNG113/140/0, and M\"OB-DAAD 27. (323-PPP-Ungarn).

\appendix

\def\theequation{\Alph{section}.\arabic{equation}}

\section{RG equations for the blocked couplings}\label{RGcoup}

\setcounter{equation}{0}

The RG equations for the blocked couplings figuring in the blocked
action (\ref{genaction}) are derived by separating the zero mode
$\varphi_0$  of the field $\varphi_k (x)$, $\varphi_k (x) = \varphi_0 
+ \eta (x)$ and expanding both sides of Eq. (\ref{WHeq}) in Taylor
series w.r.t. $\eta (x)$ terminated at the quadratic terms.
 The zeroth, first, and second order terms are
\bea
\sigma_0 &=& \int d^d x V ( \varphi_0 ) ,\nonu
\sigma_1 &=&
  \int d^d x  \left\{
      - \left\lbrack Z_{(0|1)} (x,k)
    + \varphi_0  Z_{(0|2)} (x,k) 
       \right\rbrack \Box \eta 
    + V^{(1)} (\varphi_0 ) \eta (x)
         \right\}  ,\\
\sigma_2 &=&
         \int d^d x \left\lbrack
    -          Z_{(1|1)} (k) 
           \frac{1}{2} \eta \Box \eta 
          - 
           Z_{(0|2)} (x,k)
                    \frac{1}{2}  \Box \eta^2
          + \frac{1}{2} V^{(2)} (\varphi_0 ) \eta^2 
                 \right\rbrack .\nonumber
\eea
Furthermore, we need the second derivative of the blocked action,
$S_k^{(2)} = A + B + C + {\it o}(\eta^3 )$ with the matrices $A$, $B$, and $C$
of zeroth, first, and second order of $\eta$ and having the following forms:
\begin{eqnarray}
A_{pq} &=& {\cal A} (p^2) V_d \delta_{p+q} + \chi_{pq}
\end{eqnarray}
with the diagonal part
\begin{eqnarray}
\label{cala}
   {\cal A} (p^2 ) &=& V_\kappa^{(2)} ( \varphi_0 )
          +
        Z_{(1|1)} p^2 
                 ,
\end{eqnarray}
and the off-diagonal piece
\begin{eqnarray}
  \chi_{pq} &=&
       \int d^d z 
          Z_{(0|2)} (z) (p+q)^2 
      e^{i (p+q) z} ,
\end{eqnarray}
furthermore
\begin{eqnarray}
  B_{pq}  &=&
     {\cal B}  \int dz \eta (z) e^{i(p+q) z}
\end{eqnarray}
with
$    {\cal B}  =   V^{(3)} (\varphi_0 )$,
and 
\begin{eqnarray}
  C_{pq} &=& \frac{1}{2} V^{(4)} (\varphi_0 ) \int d^d z \eta^2 
 e^{i(p+q) z} .
\end{eqnarray}
Rewriting the logarithm on the r.h.s. of Eq. (\ref{WHeq}) as
  $\ln (A+B+C+ \ldots )=
\ln A + \ln \left\lbrack 1 + A^{-1} (B+C+ \ldots )\right\rbrack$ and
expanding  the second logarithmic term in Taylor series, $A^{-1} B +
A^{-1}C - {1 \over 2} A^{-1} B A^{-1} B \ldots $, one finds the
equations
\begin{eqnarray}
\label{zeroth}
  k \partial_k \sigma_0 &=&
   - k^d \alpha_d \int \frac{d\omega_n}{\Omega_d} 
           \left( \ln A \right)_{kn, -kn} ,
          \\
\label{first}
   k \partial_k \sigma_1 &=&
   - k^d \alpha_d \int \frac{d\omega_n}{\Omega_d} 
 \left(  A^{-1} B
 \right)_{kn, -kn} ,
            \\
\label{second}
   k \partial_k \sigma_2 &=&
    - k^d \alpha_d \int \frac{d\omega_n}{\Omega_d} 
 \left(  A^{-1} C  - {1 \over 2} A^{-1} B A^{-1} B
 \right)_{kn, -kn}  .
\end{eqnarray}
 Here, 
the  matrix products of the form $(MN)_{pq}=\sum_P M_{p,-P} N_{P,q}$
should be understood over a restricted phase space, i.e., the sum over
$P$ should be restricted  to the thin momentum shell $k < |P| \le k
 +\delta k$. Such sums are performed in the continuum limit as integrals
over the interval $\lbrack k-\epsilon, k+\delta k \rbrack$ with
$\epsilon >0$ infinitesimal and the limits $\epsilon \to 0$ and
$\delta k \to 0$ are taken at the end. It has been checked that this
procedure provides a result for field independent wave function
renormalization $Z_{(1|1)}$ in IMA which is in
agreement with the  perturbative one-loop result obtained e.g. in
\cite{Col84} for $\phi^3$ theory in $d=6$ dimension.

\underbar{$\sigma_0$:}
In order to find the explicit form of Eq. (\ref{zeroth}), we write
\begin{eqnarray}
 ( \ln A )_{p, -p} &=& {\mbox {const.}} + \ln {\cal A} (p^2) +
   {1 \over V_d} ({\cal A}^{-1} \chi )_{p,-p}
   - \frac{1}{2V_d^2}   ({\cal A}^{-1} \chi {\cal A}^{-1} \chi )_{p,-p}
     + {\it o} (\chi^3) 
\end{eqnarray}
Since $\chi$ contains second powers of the momenta, its first, second,
etc. powers
generate gradient terms of the order $\partial^2$, $\partial^4$, etc.,
resp. In the second order of the gradient expansion we only have to
include the term linear in $\chi$, but its diagonal matrix element
vanishes, so that we obtain $ ( \ln A )_{p, -p} \approx  \ln {\cal A} (p^2)$
and the integro-differential  equation 
\begin{eqnarray}
\label{potbl}
   V_k (\varphi_0 )
         &=&
    V_\Lambda (\varphi_0) 
     +
  \hbar \alpha_d \int_k^\Lambda 
   d \kappa \; \kappa^{d-1}
   \ln  {\cal A} (\kappa^2 ) 
\end{eqnarray}
for the blocked potential $V_k (\varphi_0 )$.
Here $\alpha_d = \Omega_d (2\pi )^{-d}/2$, $\Omega_d = 2\pi^{d/2}/
\Gamma( d/2)$ is the entire solid 
angle in the $d$ dimensional momentum space.

\underbar{$\sigma_1$:} For the evaluation of the r.h.s. of Eq.
(\ref{first}) we need the inverse of the non-diagonal
matrix $A_{pq}$. Expanding it in powers of the off-diagonal matrix
$\chi_{pq}$, one finds
\begin{eqnarray}
\label{Ainv}
   (A^{-1} )_{pq} &=& {1 \over V_d} {\cal A}^{-1} (p^2) \delta_{p+q}
        -  {1 \over V_d^2} {\cal A}^{-1} (p^2) \chi_{pq} {\cal A}^{-1} (q^2)
  + {\it o}
        (\chi^2) .
\end{eqnarray}
The trace of the matrix product 
\begin{eqnarray}
 (A^{-1} B)_{pq} &=&
    {1 \over V_d } {\cal A}^{-1} (p^2) B_{pq} 
       - {1 \over V^2_d}  {\cal A}^{-1} (p^2) \int ' \frac{ V_d d\omega_P
      P^{d-1} dP }{ (2\pi )^d}   \chi_{p,-P}  {\cal A}^{-1} (P^2) B_{Pq}
\end{eqnarray}
can be evaluated by performing the  integral over the infinitesimal
momentum shell as discussed above.
Substituting the result in the r.h.s. of Eq. (\ref{first})
and comparing the corresponding terms on its both sides, one gets:
\begin{eqnarray}
\label{pot1}
   k \partial_k V^{(1)} (\varphi_0 ) &=& 
    - \hbar k^d \alpha_d 
       {\cal B}  {\cal A}^{-1} (k^2)  ,
             \\
\label{sig1a1}
  k \partial_k 
            Z_{(0|1)} (k) 
          &=&
    \hbar k^d \alpha_d
         {\cal B}_0  {\cal A}_0^{-2}  
            Z_{(0|2)} (k) ,
         \\
\label{sig1a2}
  k \partial_k
            Z_{(0|2)} (k) 
      &=&
    \hbar k^d \alpha_d
     \left\lbrack
  \frac{ V^{(4)} (0) }{  {\cal A}_0^2  }
     -       
 \frac{ 2{\cal B}^2  }{ {\cal A}_0^3  }
        \right\rbrack
            Z_{(0|2)} (k) .
\end{eqnarray}
Here
Eq. (\ref{pot1}) is the first derivative of Eq. (\ref{potbl}),
${\cal A}_0 = \left. {\cal A} (k^2) \right|_{\varphi_0 =0}$,
${\cal B}_0  =\left. {\cal B}  \right|_{\varphi_0 =0}$.

\underbar{$\sigma_2$:}
The trace of the first matrix on the r.h.s. of Eq. (\ref{second})
can be evaluated analogously to that of $A^{-1} B$:
\begin{eqnarray}
  -\hbar k^d \alpha_d   \int \frac{d\omega_n}{\Omega_d} 
  \frac{ V^{(4)} (\varphi_0 ) }{ {\cal A} (k^2) }
    \int d^d z \frac{1}{2}
    \left\lbrack \eta^2 (z) + Z_{(0|2)} (z) \Box \eta^2 (z)
  \right\rbrack 
.
\end{eqnarray} 
Up to the first order of $\chi_{pq}$, one can write
\begin{eqnarray}
  - \frac{1}{2} \left(   A^{-1} B A^{-1} B \right)_{pq}
     &=&
  - \frac{1}{2V_d^2} \sum_P ' \frac{ B_{p,-P} B_{Pq} }{
             {\cal A} (p^2) {\cal A} (P^2)                }
  +  \frac{1}{2V_d^3} \sum_P ' \sum_Q '
      \frac{ B_{p,-P} \chi_{P, -Q} B_{Qq} }{
             {\cal A} (p^2) {\cal A} (P^2) {\cal A} (Q^2)  }
                \nonumber\\ 
         &   &
   +    \frac{1}{2V_d^3} \sum_P ' \sum_Q '
      \frac{ \chi_{p,-Q} B_{Q, -P} B_{Pq} }{
             {\cal A} (p^2) {\cal A} (P^2) {\cal A} (Q^2)  } .
\end{eqnarray}
The second and third terms give identical contributions to the r.h.s.
of Eq. (\ref{second}),
\begin{eqnarray}
  \hbar k^d \alpha_d \frac{1}{2} \int \frac{d\omega_n}{\Omega_d} 
        \frac{ {\cal B}^2  }{ {\cal A}^3 (k^2) }
    \int d^d z  Z_{(0|2)} (z) \Box
      \eta^2
   (z) ,
\end{eqnarray}
where only the terms of second order in the gradient are retained.
The first term leads to the contribution
\begin{eqnarray}
\label{firter}
  \frac{\hbar k^d \alpha_d }{2} \int \frac{d^d p_1}{(2\pi )^d}
    \frac{ {\cal B}^2   }{ {\cal A} (k^2) 
          {\cal A} ( (kn + p_1)^2 )            } 
         \eta_{p_1} \eta_{-p_1} 
\end{eqnarray}
after performing $\sum_P '$. Expanding  the integrand in powers of $p_1^\mu$, 
 we find for the contribution (\ref{firter}):
\begin{eqnarray}
  \hbar k^d \alpha_d \frac{1}{2} \int \frac{d\omega_n}{\Omega_d}
  \left\{
 \frac{ {\cal B}_0^2 }{{\cal A}^2 (k^2) }
    \int d^d z \eta^2 (z)
  -  {\cal G}_{\rho \sigma} \int d^d z \eta (z) \partial^\rho
  \partial^\sigma \eta (z)
        \right\} ,
\end{eqnarray}
where
\begin{eqnarray}
   {\cal G}_{\rho \sigma}  &=&
      \frac{ {\cal B}_0^2 Z_{(1|1)} (k)  }{       
              {\cal A}^3 (k^2) }   
      \left\lbrack
       \frac{ 4k^2   }{       
              {\cal A} (k^2) }  Z_{(1|1)} (k)  n_\rho n_\sigma
        - g_{\rho \sigma } \right\rbrack .
\end{eqnarray}
Adding  all the contributions
on the r.h.s. of Eq. (\ref{second}), we can identify the coefficients of
the various composite operators on its both sides. 
Removing now the inhomogeneity of the couplings, we          
 find:
\begin{eqnarray}
\label{sig2a}  
  k \partial_k V^{(2)} (\varphi_0 ) &=&
    - \hbar k^d \alpha_d 
      \partial_{\varphi_0}^2 \ln {\cal A} (k^2) ,
    \\
\label{sig2c1}
   k \partial_k Z_{(1|1)} (k) 
               &=&
       \hbar k^d \alpha_d \frac{1}{d}  
               {\cal G}^\mu_\mu (0) ,
               \\
\label{sig2d}
   k  \partial_k Z_{(0|2)} (k) 
          &=&  
   \hbar k^d \alpha_d 
       \left\lbrack 
      \frac{ V^{(4)} (0) }{ {\cal A}_0^2  }
     -  \frac{ 2 {\cal B}^2  }{ {\cal A}_0^3   }
       \right\rbrack 
            Z_{(0|2)} (k) 
\end{eqnarray}
with  ${\cal G}^\mu_\mu (0) = \left. {\cal G}^\mu_\mu
\right|_{\varphi_0 =0}$.
Eq. (\ref{sig2a}) can also be obtained by differentiating both sides of Eq. 
(\ref{potbl}) w.r.t. $\varphi_0$ two times.
 Eq.
(\ref{sig2d}) is equivalent with Eq. (\ref{sig1a2}), whereas
Eq. (\ref{sig2c1}) is an independent equation.

\section{Solution of RG equations for the blocked couplings in
IMA}\label{coupIMA}

\setcounter{equation}{0}

We rewrite Eqs. (\ref{potbl}), (\ref{sig1a1}), (\ref{sig1a2}),
and (\ref{sig2c1}) in integral form and 
substitute the bare values $g_n (\Lambda )$ for  the blocked couplings
$g_n (k)$ in the integrands. In terms of the integrals $I_0$ and
$I_{nrs}$ given in Appendix \ref{integrals}, we have
\begin{eqnarray}
\label{potIMA}
       V_k (\varphi_0 ) &=& V_\Lambda  (\varphi_0 )
 + \hbar \alpha_d I_0 ,
              \\
\label{02IMA}
        Z_{(0|2)} (k)
            &=&
           Z_{(0|2)} (\Lambda)
                   -
        \hbar \alpha_d \left( g_4 (\Lambda ) I_{002} - 2I_{023}
                      \right)
         Z_{(0|2)} (\Lambda) ,
               \\
\label{01IMA}
           Z_{(0|1)} (k) 
            &=&
         Z_{(0|1)} (\Lambda) 
                   -
        \hbar \alpha_d  I_{012}
              Z_{(0|2)} (\Lambda) ,
               \\
\label{11IMA}
    Z_{(1|1)} (k) &=& Z_{(1|1)} (\Lambda ) \left\lbrack 
          1   - \hbar \alpha_d 
         \left(  
           \frac{ 4Z_{(1|1)}(\Lambda ) }{d} I_{124} 
        -  I_{023}
          \right) \right\rbrack  .
\end{eqnarray}

Let us evaluate the   elements $s_{nm}$ of the operator mixing matrix
for the $\phi^3 $ theory.

\underbar{From monomials to monomials:}
We have from Eq. (\ref{potIMA})
\begin{eqnarray}
\label{JnkIMA}
     g_\ell (k) &=& g_\ell (\Lambda ) + \hbar \alpha_d  I_0^{(\ell)} (0) ,
\end{eqnarray}
and, for the  mixing coefficients of the 
monomial operators,
\begin{eqnarray}
s_{\ell \ell'} &=& \frac{\partial g_\ell(k)}{\partial g_{\ell'}(\Lambda)}
    = \delta_{\ell \ell'} + \hbar \alpha_d
 \frac{\partial I_0^{(\ell)} (0) }{ \partial g_{\ell '} (\Lambda ) } .
\end{eqnarray}
The  partial derivatives on the r.h.s. of Eq. (\ref{JnkIMA})
can be evaluated by taking the  partial derivatives of $I_0$
(given in Appendix \ref{integrals}) w.r.t. $\varphi_0$  at $\varphi_0=0$
and then differentiating them w.r.t. the appropriate $g_{\ell'} (\Lambda)$,
before the integration over $\kappa$ is performed.
Finally, the initial values of the couplings specifying the $\phi^3$
theory should be inserted.

In Table \ref{imas} we summarize the non-trivial operator mixing
coefficients in terms of the integrals given in Appendix \ref{integrals}
and their limiting values for $k=0$, $\Lambda^2 \gg m^2$
after carrying out the loop integral in IMA.
The functions $f_a$, $a=0,2,4$ in the last column are defined as
\begin{eqnarray}
 f_0 (\Lambda ) &=&
             \left\lbrack
           \ln \left( \frac{\Lambda^2}{m^2} +1 \right)
           - \frac{3}{2} \right\rbrack
               ,
          \nonumber\\
 f_2 (\Lambda ) &=&
 \left\lbrack
               m^2  \ln \left( \frac{\Lambda^2}{m^2} +1 \right)
              - \frac{1}{2} \Lambda^2 - \frac{1}{2} m^2
               \right\rbrack ,
          \nonumber\\
f_4 (\Lambda ) &=&\left\lbrack
          m^4 \ln \left( \frac{\Lambda^2}{m^2} +1 \right)
         + \frac{1}{2} \Lambda^4 - \Lambda^2 m^2
           \right\rbrack,\nonu
f_6 (\Lambda ) &=&       -
     \frac{1}{3} \Lambda^6
   + \frac{1}{2} \Lambda^4 m^2 - \Lambda^2 m^4
   + m^6 \ln \left( \frac{\Lambda^2}{m^2} +1 \right) .
\end{eqnarray}

The theory is renormalizable and we see that  monomial
operators $\varphi^{\ell'}$ of dimension not greater than the dimension
of the monomial $\varphi^\ell$ are the only ones admixing to the blocked
operator  $\{\varphi^\ell\}_{k\to0}$
with UV divergent coefficients. Were the theory non-renormalizable,
e.g. a $\phi^4$ term with non-vanishing coupling $g_4(\Lambda)\neq0$
included, then also the admixture of higher dimensional operators
would occur with UV divergent coefficients.

These results can
be compared with the one-loop perturbative result given in
\cite{Col84} (p.145, (6.2.11)) for the renormalized operator 
$\frac{1}{2} \lbrack \phi^2 \rbrack_R$, replacing $2 / (d-6)$
 by $- \ln (\Lambda^2/m^2 )$ and using $\alpha_6
= 1/(128\pi^3)$ (in our notations):
\begin{eqnarray}
\label{Colphi2}
  \frac{1}{2} \lbrack \phi^2 \rbrack_R  &=&
     M_1 (\phi ) \left( Z^{-1}\right)_{12}
  +  M_2 (\phi )  \left( Z^{-1}\right)_{22}
  +  D_{(0|1)} (\phi ) \left( Z^{-1}\right)_{(0|1)2} +
        \ldots
\end{eqnarray}
where $\ldots$ stands for higher order terms and
\begin{eqnarray}
     \left( Z^{-1}\right)_{12} = - \hbar \frac{\lambda m^2}{128 \pi^3}
           \ln \frac{ \Lambda^2}{m^2} ,
       \qquad
      \left( Z^{-1}\right)_{22} =
      1 - \hbar \frac{\lambda^2}{128\pi^3} 
  \ln \frac{ \Lambda^2}{m^2} ,
       \qquad
    \left( Z^{-1}\right)_{(0|1)2}  &=&
     \hbar \frac{ \lambda}{6 \cdot 128\pi^3}
 \ln \frac{ \Lambda^2}{m^2}.
\end{eqnarray}
 As far
as the mixing of monomials is considered,
 this illustrates that the operator
mixing matrix $s_{nm}$ in the IMA in the limit $k \to 0$ and $\Lambda^2
\gg m^2$ satisfies the relation $s_{\ell \ell'} = Z_{\ell \ell'}$, where
$Z_{nm}$ is the one-loop perturbative operator mixing matrix.

\underbar{From monomials to derivative operators:}
Eq. (\ref{JnkIMA}) obtained from  (\ref{potIMA})
can also be used to determine the admixture of monomials to the
blocked derivative operators in the IMA. Since $I_0^{(n)} (0)$ depends
only on $z =Z_{(1|1)} (\Lambda )$ and not on the couplings of the
other derivative operators, one obtains $s_{\ell(r|s) } (k) =0$ for
$(r|s) \neq (1|1)$,   and
\begin{eqnarray}
  s_{\ell(1|1)} (k) & =& \hbar \alpha_d \frac{\partial I_0^{(n)} (0)
    }{\partial z} .
\end{eqnarray}
The  first few non-vanishing operator mixing coefficients and their
limiting values for  $k=0$, $\Lambda^2 \gg m^2$ are collected in
Table \ref{imask}. Thus, we can write for the  blocked operator (with
constant argument!):
$\{D_{(1|1)}(\varphi_0)\}_{k}=\sum_{n=0}^\infty
s_{n(1|1)}(k)\varphi_0^n /n!$.
Again, the contributions of the  derivative operators to the
blocked derivative operator $ \{ D_{(1|1)}(\varphi_k)  \}_{k} $
can only be seen if the latter is
considered at the general argument $\varphi_k (x)$.

\underbar{From derivative operators to monomials and derivative
operators:} Differentiating the solutions
(\ref{11IMA}), (\ref{02IMA}), and  (\ref{01IMA})
w.r.t. the various bare couplings the following
additional (non-vanishing) operator mixing coefficients and
their limiting values for $k=0$, $\Lambda^2 \gg m^2$ are shown
in Table \ref{imasn}.

The result obtained for $Z_{(1|1)} (k=0)$ in the IMA is in agreement
with the one-loop perturbative result for $Z_\phi$: 
\begin{eqnarray}
\label{dim6Zc}
  Z_\phi &=& 1 + \frac{\hbar \lambda^2}{ 12 \cdot 64
    \pi^3}
 \left(
        \ln \frac{\Lambda^2}{m^2} 
        - \frac{5}{6} \right) 
\end{eqnarray}
(see \cite{Col84}, p.58)
as expected, since the correspondence $Z_{(1|1)} (k=0) = Z_\phi$ 
 should hold due to 
\begin{eqnarray}
\int d^d x \frac{1}{2} Z_\phi (\partial_\mu \phi )^2 = 
 \int d^d x \frac{1}{2} Z_\phi ( 2 D_{(1|1)} - D_{(0|2)} )
         = \int d^d x Z_{(1|1)}  D_{(1|1)}.
\end{eqnarray}
Using $Z_{(0|2)} (\Lambda ) = -{1 \over 2}$ (owing to
Eq. (\ref{dfidfi})) and  $Z_{(0|1)} (\Lambda ) =0$,
a comparison of $s_{(0|1)2}$ with the corresponding one-loop
perturbative result (see relation (\ref{Colphi2}) taken from
\cite{Col84})
 gives $s_{(0|1)2} = 3 Z_{ (0|1)2}$. Since $s_{(0|1)2}$ obtained above
agrees with our perturbative one-loop result
in Appendix \ref{per1loop}, so that both results  contradict to 
(\ref{Colphi2}),
there ought to be a misprint in \cite{Col84}.

\section{Connection matrix}\label{senmat}

\setcounter{equation}{0}

First, we read off the beta-functions from the right hand sides of
Eqs.  (\ref{potbl}), (\ref{sig1a1}), (\ref{sig1a2}),
and (\ref{sig2c1}). The beta-functions for the
couplings of the monomials are given by $\beta_\ell (k) = \hbar \alpha_6
k^6 \left. \partial_{\varphi_0}^\ell \ln {\cal A} (k^2 )
\right|_{\varphi_0 =0} $, whereas the other beta-functions can be 
read off directly. With the notations $m^2(k)=V^{(2)}(0)$,
$\lambda(k)=V^{(3)}(0)$, $g(k)=V^{(4)}(0)$, $g_\ell(k)=V^{(\ell>4)}(0)$, and
$G=\left\lbrack Z_{(1|1)} (k) k^2 + m^2 (k) \right\rbrack^{-1}$,
the explicit forms of the beta-functions are given in Table \ref{betat}.

The elements of the connection matrix, shown in Table \ref{connt}
are obtained by differentiation
according to the second equation of (\ref{Yijsij}).

\section{Perturbative one-loop results for operator mixing}\label{per1loop}

\setcounter{equation}{0}

Here we determine the operator mixing matrix at one-loop order
perturbatively for $\lambda \phi^3$ theory in dimension $d=6$.
Let us decompose the action $S$ into the free part
\begin{eqnarray} 
 S_0 &=&
  \int d^d x \left\lbrack
       - \frac{1}{2} Z_R \phi \Box \phi + \frac{1}{2} m_R^2 \phi^2
               \right\rbrack ,
\end{eqnarray}
the interaction part
\begin{eqnarray}
   S_I &=&
 \int d^d x \left\lbrack
      -\frac{1}{2} g_{(1|1)R} (x) \phi \Box \phi
      + \frac{1}{2} g_{2R}(x) \phi^2
      + \sum_{n \neq (1|1), 2 } G_{nR} (x) O_n (\phi (x) )
        \right\rbrack ,
\end{eqnarray}
and the counterterms
\begin{eqnarray}
  S_{c.t.} &=& 
       \int d^d x
        \sum_n c_n ( g_{mR} ) O_n ( \phi (x) ) .
\end{eqnarray} 
Here the same base operators are taken into account as in the blocked
action (\ref{genaction}).
The one-loop effective action is given by 
\begin{eqnarray}
   \Gamma_1 &=&
      \frac{\hbar}{2} {\mbox { Tr }} \ln 
       \left( 1_{pq} + G_R (p^2 ) (S_I^{(2)})_{-p \; q}
  \right) 
              \nonumber\\
         &=&
    - \frac{\hbar }{2} \sum_{v=1}^\infty 
        \frac{ (-1)^v}{v} \int_{p_1} \cdots \int_{p_v} 
     G_R ( p_1^2 )  (S_I^{(2)})_{-p_1 \; p_2}
             \cdots
     G_R ( p_v^2 )  (S_I^{(2)})_{-p_v \; p_1}
\end{eqnarray}
with $G_R (p^2) =  (Z_R p^2 + m_R^2 )^{-1}$ and
the second functional derivative of the interaction piece of the action:
\begin{eqnarray}
   ( S_I^{(2)} )_{pq} &=&
         \int dx e^{i(p+q)x } 
          \left\lbrack g_{2R} (x) + g_{(1|1)R} (x) \frac{1}{2} (p^2 +
            q^2 )
                   \right.
                \nonumber\\
         &   &
      \left.
         + \sum_{\ell=3}^\infty \left( g_{\ell R} + g_{\ell R} (x) \right)
             \frac{1}{(\ell -2)!} \phi^{\ell -2} (x)
         + \left( Z_{(0|2)R} + g_{(0|2)R} (x) \right)
               ( p+q)^2 
              \right\rbrack .
\end{eqnarray}
In the second order of the derivative expansion, the counterterms are
defined through the relations:
\bea
   - \left. \frac{ \delta \Gamma_1}{\delta \varphi (p)}
       \right|_{\phi, g_{{\tilde n}R} =0} &=&
  \left( c_1 +  c_{(0|1)} p^2 + {\it o} (p^4) \right)
       (2 \pi )^d \delta (p) ,\nonu
   -  \left.
 \frac{ \delta^2 \Gamma_1 }{ \delta \varphi (p_1 ) \delta \varphi (p_2 )
     } \right|_{\phi, g_{{\tilde n}R} =0}  
               &=&
   \left( c_2 +  c_{(1|1)} p_1^2 + {\it o} (p_1^4) \right)
       (2 \pi )^d \delta (p_1 + p_2) ,\\
      -  \left.
 \frac{ \delta^s \Gamma_1 }{ \delta \varphi (p_1 ) \cdots \delta \varphi (p_s )
     } \right|_{\phi, g_{{\tilde n}R} =0, p_i^2 =0}
             &=&  
            c_s  
       (2 \pi )^d \delta (p_1 + \ldots + p_s) ,  \qquad
             ( s \ge 3 )\nonu
  - Z_{(0|2)R} \left.
 \frac{ \delta^3 \Gamma_1 }{ \delta \varphi (p_1 )  \delta \varphi (p_2 )
          \delta g_{(0|2)R} (x)
     } \right|_{\phi, g_{{\tilde n}R} =0}
             &=&  
            c_{(0|2)} 
      (p_1 + p_2 )^2 e^{i(p_1 + p_2 ) x }  .
               \nonumber
\eea
A straightforward evaluation results in the expressions for
the counterterms given in Table \ref{counterm}.

For the elements of the operator mixing matrix (\ref{defzmn})
we obtain, with $Z_R \equiv Z_{(1|1)R}  =1$, $Z_{(0|2)R} = - {1 \over
  2}$, $g_{nR} =0$ for $n \neq 2, (1|1),  (0|2), 3$,
the  expressions shown in Table \ref{permixmat}.

\section{Particular integrals}\label{integrals}

\setcounter{equation}{0}

In order to obtain the explicit forms of the solutions of
Eqs. (\ref{potIMA}), (\ref{11IMA}), (\ref{01IMA}), and (\ref{02IMA})
the following integrals are needed:
\begin{eqnarray}
  I_0 &=& \int_k^\Lambda dpp^{d-1} \ln \left(zp^2 +
  V_\Lambda^{(2)} (\varphi_0 )    \right) ,
            \nonumber\\
  I_{nrs} &=& \lambda_\Lambda^r\int_k^\Lambda dpp^{d-1} \; p^{2n} G_\Lambda^s
  =
 \frac{\lambda_\Lambda^r}{2z^s}
       \sum_{i=0}^{j+n-1}\pmatrix{j+n-1\cr i}
         \left( - \frac{ m^2_\Lambda}{z} \right)^{j+n-1-i}
 \int_{k^2}^{\Lambda^2} du
 \left( u + \frac{m^2_\Lambda}{z} \right)^{i-s} 
\end{eqnarray}
for even dimensions $d=2j$, where
$G_\Lambda =\left\lbrack zp^2 + m^2_\Lambda
 \right\rbrack^{-1}$.
For dimensions $d=6$ we obtain:
\begin{eqnarray}
  I_0 &=& \frac{1}{6} \left\lbrack 
      u^3 \ln \left( z  u + V_\Lambda^{(2)} (\varphi_0 ) \right)
    - \frac{1}{3} ( u+c)^3 + \frac{3}{2} c ( u+c)^2 - 3c^2 u 
        + c^3 \ln (u+c) \right\rbrack^{u=\Lambda^2}_{u=k^2} 
\end{eqnarray}
with $ c=  V_\Lambda^{(2)} (\varphi_0 ) /z$.
The evaluation of the integrals $I_{nrs}$ is straightforward.

\newpage
\mediumtext

\begin{table}[tp]
\caption{Comparison of various notions in the renormalization group (RG)
method and the renormalized perturbation expansion (RPE)}\label{parallel}
\begin{tabular}{|c|c|}
     \tableline
       &    \\
  RG & RPE  \\
        &    \\
   \tableline
        &     \\
   coupling at UV scale:
     $ g_n ( \Lambda)$ &  bare coupling: $g_n$ \\
  blocked coupling:
                      $ { g}_n (k)$ &   renormalized coupling: $g_{nR}$ \\
          &   \\
    \tableline
        &       \\
  operator mixing matrix:  &  operator mixing matrix:  \\
             $s_{nm} (k,\Lambda)$     &
$ Z_{nm} $ \\
     defined as           &        defined via \\
$ s_{nm}
= {\partial g_{n} (k) \over\partial g_m (\Lambda)}
  $  &
  $ \left( Z^{-1} \right)_{mn}  =
 {\partial   g_{m}  \over \partial g_{nR}}$
 \\
               &   \\
  \tableline
             &\\
 operator at UV scale: $O_{\tilde n} (\phi )$
               &
  bare operator: $O_{\tilde n} (\phi )$
              \\
       &   \\
 \tableline
         &  \\
                          blocked operator:  &     not used  \\
 $ \{ O_{\tilde n}  (\varphi_k ) \}_k
            =
         \left.
 {\delta S_k \over \delta
          G_{\tilde n} ( \Lambda )}
     \right|_{g_{\tilde m} ( \Lambda) =0}   $  &
        \\
   operator mixing: & \\
    $ \{ O_{\tilde n}  (\varphi_k ) \}_k  =
    \sum_m  O_{\tilde m} (\varphi_k ) s_{mn} (k)$ &
   \\
         &  \\
          \tableline
         &   \\
   operator obtained by   &  renormalized operator:\\
    inverse blocking        &                        \\
   (if exists):          &                       \\
   $ \lbrack O_{\tilde n} (\phi ) \rbrack_k
         = \sum_m  O_{\tilde m} (\phi  ) ( s^{-1})_{mn} (k,\Lambda)$
               &
 $ \lbrack O_{\tilde n} (\phi  ) \rbrack_R = \sum_m  O_{\tilde m} (
 \phi ) (Z^{-1})_{mn} $              \\
           &   \\
        \tableline
\end{tabular}
\end{table}

\begin{table}[tp]
\caption{Orders of magnitude of the operator mixing coefficients
$s_{nm}$ in IMA for $\lambda \phi^3$ theory for $k=0$, $\Lambda^2\gg m^2$.
$E (\Lambda )$ indicates a constant plus
logarithmically divergent one-loop contributions; $M_n =\varphi^n/n!$;
$d_n$, $d_m$ stand for the momentum dimensions of the
operators $\{ O_{\tilde n} \}$, $O_{\tilde m}$, resp.}\label{YijIMA}
\begin{tabular}{|c|c||c||c||c|c||c|c|c||c||}
   \tableline 
   & $d_m$ &0&2& \multicolumn{2}{l||}{4} &
                \multicolumn{3}{l||}{6} & 8 \\
    \tableline
   $d_n$ & $ \{ O_{\tilde n} \}$ $\; \setminus \; $ $O_{\tilde m}$
           &$ M_0$ &$M_1$& $M_2$ & $D_{(0|1)}$ &
           $M_3$ & $D_{(0|2)}$ & $D_{(1|1)}$ &
           $M_4$      \\
      \tableline
      \tableline
  0& $\{ M_0 \}$ & 1 & 0& $\Lambda^4$ & 0 & 0 & 0 & $\Lambda^6$ & 0  \\
      \tableline
      \tableline
 2 & $\{ M_1 \}$& 0& 1 & $\Lambda^2$ & 0 & $\Lambda^4$ & 0 &
 $\Lambda^4$ & 
 0  \\      
     \tableline
      \tableline
 4 & $\{ M_2 \}$&0 & 0 & $ E( \Lambda )$
      & 0 & $\Lambda^2$ &    $0$  & $\Lambda^2$
 & $\Lambda^4$     \\
       \cline{2-10}
   & $\{ D_{(0|1)} \}$ & 0&0&$\ln \Lambda$& 1&
   $\Lambda^2$&$\Lambda^2$&
      $\Lambda^2$ &0  \\
       \tableline
       \tableline
 6 & $\{ M_3 \} $ & 0 & 0 & $\Lambda^0 $ & $0$ &
      $E (\Lambda ) $ & $0$ &  $\ln \Lambda $ &
      $\Lambda^2$   \\
        \cline{2-10}
   & $\{ D_{(0|2)} \}$ & 0 &0&$\Lambda^0$ & $0$ & $\ln \Lambda$ &
        $ E ( \Lambda )$ & $\ln \Lambda$ & $\Lambda^2$  \\
        \cline{2-10}
   & $\{ D_{(1|1)} \}$ & $0$ & $0$ & $\Lambda^0$ &
  $0$ &  $\ln \Lambda $ &  $0 $ &
              $ E (\Lambda ) $ & $0$   \\
         \tableline 
         \tableline
   8 & $\{ M_4 \}$ & 0  & 0  & $\Lambda^0$  & 0  &$\Lambda^0$  & 
         $0$  & $\Lambda^0$  & $E( \Lambda )$    \\
 \tableline
\end{tabular}
\end{table}

\begin{table}[tp]
\caption{Elements of the connection matrix $\gamma_{nm} (k)$
  for a theory with polynomial potential; the non-trivial matrix
  elements are indicated by $\times$, $d_n$, $d_m$ denote the momentum
  dimensions of the base operators $O_{\tilde n} (\phi )$,  $O_{\tilde
    m}(\phi )$,
  resp.
 }\label{sensim}
\begin{tabular}{|c|c||c||c||c|c||c|c|c||c||}
   \tableline
    & $d_m$ &0&2& \multicolumn{2}{l||}{4} &
                \multicolumn{3}{l||}{6} & 8 \\
    \tableline
   $d_n$& $n$ $\; \setminus \; $ $m$ &$ 0$ &$1$& $2$ & ${(0|1)}$ &
           $3$ & ${(0|2)}$ & ${(1|1)}$ &
           $4$      \\
      \tableline
      \tableline
  0& $0 $ & 0 & 0& $\times$ & 0 & 0 & 0 & $\times$ & 0  \\
      \tableline
      \tableline
 2 & $ 1$& 0& 0 & $\times$ & 0 & $\times$ & 0 & $\times$ & 0     \\
     \tableline
      \tableline
 4 & $2 $& $0 $ & $0$ & $\times$
      & $0$ & $\times$ &    $0$  & $\times$
 & $\times$  \\
       \cline{2-10}
   & ${(0|1)} $ & 0&0&$\times$& 0&$\times$ & $\times$
    & $\times$   & 0 \\
       \tableline
       \tableline
 6 & $3  $ & 0 & $0$ & $\times $ & $0$ &
      $\times$ & $0 $ &  $\times $ &
      $\times$   \\
        \cline{2-10}
   & ${(0|2)} $ & 0 &0&$\times$ & $0$ & $\times$ &
        $\times $ &  $\times $  & $\times$ \\
        \cline{2-10}
    & ${(1|1)} $ & 0 &0
   & $\times$ & 0   & $\times$ & 0
   & $\times$  & 0
              \\
         \tableline
         \tableline
   8 & $4 $ & 0  & 0  & $\times$  & 0  &$\times$  &
         $0$  &  $\times$  & $\times $
          \\
 \tableline
\end{tabular}
\end{table}

\begin{table}[tp]
\caption{The mixing matrix elements $s^{IMA}_{mn}(k)$.\label{imas}}
\begin{tabular}{|c|c|c|}
\tableline
$m,n$ & integral form & after integration \\
\tableline
$0,2$&$\hbar\alpha_6\partial_{m^2} I_0(0)=\hbar\alpha_6I_{001}$
&$\hbar\alpha_6f_4(\Lambda )/2$\\
\tableline
$1,2$&$\hbar\alpha_6\partial_{m^2}I_0^{(1)}(0)=-\hbar\alpha_6I_{012}$
&$\hbar\alpha_6\lambda f_2(\Lambda )$\\
\tableline
$2,2$&$1+\hbar\alpha_6\partial_{m^2}I_0^{(2)}(0)=1+2\hbar\alpha_6I_{023}$
&$1+\hbar\alpha_6\lambda^2f_0(\Lambda )$\\
\tableline
$3,2$&$\hbar\alpha_6\partial_{m^2}I_0^{(3)}(0)=-6\hbar\alpha_6I_{034}$
&$-\hbar\alpha_6\lambda^3/m^2$\\
\tableline
$4,2$&$\hbar\alpha_6\partial_{m^2}I_0^{(4)}(0)=24\hbar\alpha_6I_{045}$
&$\hbar\alpha_6\lambda^4/m^4$\\
\tableline
$1,3$&$\hbar\alpha_6\partial_\lambda I_0^{(1)}(0)=\hbar\alpha_6I_{001}$
&$\hbar\alpha_6f_4(\Lambda)/2$\\
\tableline
$2,3$&$\hbar\alpha_6\partial_\lambda I_0^{(2)}(0)=-2\hbar\alpha_6I_{012}$
&$2\hbar\alpha_6\lambda f_2(\Lambda)$\\
\tableline
$3,3$&$1+\hbar\alpha_6\partial_\lambda I_0^{(3)}(0)=1+6\hbar\alpha_6I_{023}$
&$1+3\hbar\alpha_6\lambda^2f_0(\Lambda)$\\
\tableline
$4,3$&$\hbar\alpha_6\partial_\lambda I_0^{(4)}(0)=-24\hbar\alpha_6I_{034}$
&$-4\hbar\alpha_6\lambda^3/m^2$\\
\tableline
$2,4$&$\hbar\alpha_6\partial_gI_0^{(2)}(0)=\hbar\alpha_6I_{001}$
&$\hbar\alpha_6f_4(\Lambda)/2$\\
\tableline
$3,4$&$\hbar\alpha_6\partial_gI_0^{(3)}(0)=-3\hbar\alpha_6I_{012}$
&$3\hbar\alpha_6\lambda f_2(\Lambda)$\\
\tableline
$4,4$&$1+\hbar\alpha_6\partial_gI_0^{(4)}(0)=1+12\hbar\alpha_6I_{023}$
&$1+6\hbar\alpha_6\lambda^2f_0(\Lambda)$\\
\tableline
\end{tabular}
\end{table}

\begin{table}[tp]
\caption{The mixing matrix elements
$s_{\ell(m|n)}^{IMA}(k)/\hbar\alpha_6$.\label{imask}}
\begin{tabular}{|c|c|c|}
\tableline
$\ell(m|n)$ & integral form & after integration \\
\tableline
$0(1|1)$&$I_{101}$&$-f_6(\Lambda)/2$\\
\tableline
$1(1|1)$&$-I_{112}$&$-\lambda[3f_4(\Lambda)-\Lambda^4+\Lambda^2m^2-m^4]/6$\\
\tableline
$2(1|1)$&$2I_{123}$&$-\lambda^2[6f_2(\Lambda)+\Lambda^2-2m^2]$\\
\tableline
$3(1|1)$&$-6I_{134}$&$-\lambda^3[3f_0(\Lambda)-1]$\\
\tableline
$4(1|1)$&$24I_{145}$&$3\lambda^4/m^2$\\
\tableline
\end{tabular}
\end{table}

\begin{table}[tp]
\caption{The mixing matrix elements $s_{index}^{IMA}(k)$.\label{imasn}}
\begin{tabular}{|c|c|c|}
\tableline
index & integral form & after integration \\
\tableline
$(1|1)2$&$\hbar\alpha_6(8Z_{(1|1)}^2 I_{125}
-9Z_{(1|1)}I_{024})/3$&$-\hbar\alpha_6\lambda^2/6m^2$\\
\tableline
$(1|1)3$&$-\hbar\alpha_6(4Z_{(1|1)}^2I_{114}-6Z_{(1|1)}I_{013})/3$
&$\hbar\alpha_6\lambda[3f_0(\Lambda)+2]/9$\\
\tableline
$(1|1)(1|1)$&$1-\hbar\alpha_6(13Z_{(1|1)}I_{124}
-3I_{023}-8Z_{(1|1)}^2I_{225})/3$
&$1-\hbar\alpha_6\lambda^2[6f_0(\Lambda)+1]/18$\\
\tableline
$(0|2)2$&$-6\hbar\alpha_6Z_{(0|2)}(\Lambda)I_{024}$
&$-\hbar\alpha_6\lambda^2Z_{(0|2)}(\Lambda)/m^2$\\
\tableline
$(0|2)3$&$4\hbar\alpha_6Z_{(0|2)}(\Lambda)I_{013}$
&$2\hbar\alpha_6\lambda Z_{(0|2)}(\Lambda) f_0(\Lambda)$\\
\tableline
$(0|2)4$&$-\hbar\alpha_6Z_{(0|2)}(\Lambda)I_{002}$
&$\hbar\alpha_6Z_{(0|2)}(\Lambda)f_2(\Lambda)$\\
\tableline
$(0|2)(0|2)$&$1+2\hbar\alpha_6I_{023}$
&$1+\hbar\alpha_6\lambda^2f_0(\Lambda)$\\
\tableline
$(0|2)(1|1)$&$-6\hbar\alpha_6Z_{(0|2)}(\Lambda)I_{124}$
&$-3\hbar\alpha_6\lambda^2Z_{(0|2)}(\Lambda)[3f_0(\Lambda)+2]/3$\\
\tableline
$(0|1)2$&$2\hbar\alpha_6Z_{(0|2)}(\Lambda)I_{013}$
&$\hbar\alpha_6\lambda Z_{(0|2)}(\Lambda)f_0(\Lambda)$\\
\tableline
$(0|1)3$&$-\hbar\alpha_6Z_{(0|2)}(\Lambda)I_{002}$
&$\hbar\alpha_6Z_{(0|2)}(\Lambda)f_2(\Lambda)$\\
\tableline
$(0|1)(0|1)$&$1$&$1$\\
\tableline
$(0|1)(0|2)$&$-\hbar\alpha_6I_{012}$
&$\hbar\alpha_6\lambda f_2(\Lambda)$\\
\tableline
$(0|1)(1|1)$&$2\hbar\alpha_6Z_{(0|2)}(\Lambda)I_{113}$
&$-\hbar\alpha_6\lambda Z_{(0|2)}(\Lambda)[6f_2(\Lambda)+\Lambda^2-2m^2]/2$\\
\tableline
\end{tabular}
\end{table}

\begin{table}[tp]
\caption{The beta-functions
 $\beta_{index}(k)/\hbar\alpha_6k^6$.\label{betat}}
\begin{tabular}{|c|c|}
\tableline
index & beta-function \\
\tableline
$0$&$-\ln G$\\
\tableline
$1$&$G\lambda$\\
\tableline
$2$&$G(g-G\lambda^2)$\\
\tableline
$(0|1)$&$G^2\lambda Z_{(0|2)}$\\
\tableline
$3$&$G(g_5-3gG\lambda+2G^2\lambda^3)$\\
\tableline
$(0|2)$&$G^2Z_{(0|2)}(g-2G\lambda)$\\
\tableline
$(1|1)$&$G^3 \lambda^2 Z_{(1|1)} (2k^2GZ_{(1|1)}-3)/3$\\
\tableline
$4$&$G(g_6-4g_5G\lambda-3g^2G+12gG^2\lambda^2-6G^3\lambda^4)$\\
\tableline
\end{tabular}
\end{table}

\begin{table}[tp]
\caption{The non-vanishing connection matrix elements
$k\gamma_{index}(k)/\hbar\alpha_6k^6$.\label{connt}}
\begin{tabular}{|c|c|}
\tableline
index & connection matrix \\
\tableline
$02$&$G$\\
\tableline
$0(1|1)$&$k^2G$\\
\tableline
$12$&$-G^2\lambda$\\
\tableline
$13$&$G$\\
\tableline
$1(1|1)$&$-k^2G^2\lambda$\\
\tableline
$22$&$G^2(-g+2G\lambda^2)$\\
\tableline
$23$&$-2G^2\lambda$\\
\tableline
$2(1|1)$&$k^2G^2(-g+2G\lambda^2)$\\
\tableline
$24$&$G$\\
\tableline
$(0|1)2$&$-2G^3\lambda Z_{(0|2)}$\\
\tableline
$(0|1)3$&$G^2Z_{(0|2)}$\\
\tableline
$(0|1)$&$G^2\lambda$\\
\tableline
$(0|1)(1|1)$&$-2k^2G^3\lambda Z_{(0|2)}$\\
\tableline
$32$&$-G^2(g_5-6gG\lambda+6G^2\lambda^3)$\\
\tableline
$33$&$-G^2(3g-6G\lambda^2)$\\
\tableline
$3(1|1)$&$-k^2G^2(g_5-6gG\lambda+6G^2\lambda^3)$\\
\tableline
$34$&$-3G^2\lambda$\\
\tableline
$(0|2)2$&$2G^3Z_{(0|2)}(g+3G\lambda^2)$\\
\tableline
$(0|2)3$&$-4G^3\lambda Z_{(0|2)}$\\
\tableline
$(0|2)(0|2)$&$G^2(g-2G\lambda^2)$\\
\tableline
$(0|2)(1|1)$&$-2k^2G^3 Z_{(0|2)}(g-3G\lambda^2)$\\
\tableline
$(0|2)4$&$G^2Z_{(0|2)}$\\
\tableline
$(1|1)2$&$G^4\lambda^2Z_{(1|1)}(9-8k^2GZ_{(1|1)})/3$\\
\tableline
$(1|1)3$&$G^3\lambda Z_{(1|1)}(4k^2GZ_{(1|1)}-6)/3$\\
\tableline
$(1|1)(1|1)$&$G^3 \lambda^2(12k^2GZ_{(1|1)}-3-8k^4G^2Z_{(1|1)}^2)/3$\\
\tableline
$42$&$G^2(-g_6+8g_5G\lambda+6g^2G-36gG^2\lambda^2+24G^3\lambda^4)$\\
\tableline
$43$&$-4G^2(g_5-6gG\lambda+6G^2\lambda^3)$\\
\tableline
$4(1|1)$&$-k^2G^2(g_6-8\lambda g_5G\lambda-6g^2G+36gG^2\lambda^2
-24G^3\lambda^4)$\\
\tableline
$44$&$-6G^2(g-2G\lambda^2)$\\
\tableline
\end{tabular}
\end{table}

\begin{table}[tp]
\caption{The counterterms expressed in terms of the integrals 
with $\lambda = g_{3R}$. \label{counterm}}
\begin{tabular}{|c|c|}
\tableline
index & $c_{index}/\hbar \alpha_d$ \\
\tableline
$02$&$G$\\
\tableline
$ 1$ &$ -  I_{011}$ \\
\tableline
$ 2$ &$ I_{022}  - g_{4R} I_{001}$ \\
\tableline
$ (0|1)$& $Z_{(0|2)R} I_{012}$ \\
\tableline
 $ 3$ &$ 3g_{4R} I_{012} - 2 I_{033}$ \\
\tableline
$  (0|2)$ &$\left( g_{4R} I_{002} -  2I_{023} \right) Z_{(0|2)R}$ \\
\tableline      
$(1|1)$ &$  \frac{4Z^2_R }{d} I_{124} - Z_R I_{023}$ \\
\tableline
 $ 4$ &$3 g_{4R}^2 I_{002} - 12 g_{4R} I_{023} + 6 I_{044}$\\
\tableline
\end{tabular}
\end{table}

\begin{table}[tp]
\caption{The operator mixing matrix $(Z^{-1})_{index}$
.\label{permixmat}}
\begin{tabular}{|c|c|}
\tableline
index & mixing matrix \\
\tableline
$10$, $11$ & $1 $\\
\tableline
$12$ &$\hbar \alpha_d  I_{012} $ \\
\tableline
$1(0|1)$  & $0$ \\
\tableline
$13$ &$ -\hbar \alpha_d  I_{001}$ \\
\tableline
$1(0|2)$ &$ 0$ \\
\tableline
$1(1|1)$  &$ \hbar \alpha_d I_{112}$ \\
\tableline
$14$  & $0 $\\
\tableline
\tableline              
$20 $, $21$ & $0 $\\
\tableline
$22$ & $1 + \hbar \alpha_d \left(
       g_{4R} I_{002} - 2 I_{023} \right)$ \\
\tableline
$2(0|1)$ &  $0$ \\
\tableline
$23$ &  $ 2 \hbar \alpha_d  I_{012}$ \\
\tableline
$2(0|2)$ & $0$  \\ 
\tableline
$2(1|1)$  &$ \hbar \alpha_d \left(
         g_{4R} I_{102} - 2  I_{123} \right)$ \\ 
\tableline
$24$  &$ -\hbar \alpha_d  I_{001}$ \\
\tableline
\tableline
$(0|1)0$,  $(0|1)1$ & $0$ \\
\tableline
$(0|1)2$ &$-2 \hbar \alpha_d  I_{013} Z_{(0|2)R}$ \\
\tableline        
$(0|1)(0|1)$  & $1$ \\
\tableline
$(0|1)3$ & $\hbar \alpha_d  I_{002}  Z_{(0|2)R}$ \\
\tableline
$(0|1)(0|2)$  &$\hbar \alpha_d  I_{012}$ \\
\tableline
$(0|1)(1|1)$  &$ - 2  \hbar \alpha_d  I_{113}   Z_{(0|2)R}$\\
\tableline
$(0|1)4$  & $0$ \\
\tableline
\tableline
$30$, $31$ & $0$\\
\tableline
$32$ &$6 \hbar \alpha_d \left(
      I_{034} - g_{4R} I_{013} \right)$ \\
\tableline
$3(0|1)$  &   $0$ \\ 
\tableline
$33$ & $ 1 
      + 3 \hbar \alpha_d \left( 
        g_{4R} I_{002} - 2 I_{023} \right)$ \\
\tableline
$3(0|2)$ & $0$ \\
\tableline
$3(1|1)$  &$6\hbar \alpha_d \left( 
        I_{134} - g_{4R} I_{113} \right)$ \\ 
\tableline
$34$  &$ 3\hbar \alpha_d  I_{012}$ \\
\tableline
\tableline
$(0|2)0$,  $(0|2)1$ & $0$ \\
\tableline
$(0|2)2$ &$\hbar \alpha_d \left( 6 I_{024} 
          - 2 g_{4R} I_{003} \right) Z_{(0|2)R}$ \\
\tableline
$(0|2)(0|1) $ & $0 $ \\
\tableline
$(0|2)3$ &$-4 \hbar \alpha_d  I_{013}  Z_{(0|2)R}$ \\
\tableline
$(0|2)(0|2) $ &$ 1 +  \hbar \alpha_d 
         \left( g_{4R} I_{002} - 2 I_{023} \right)$ \\ 
\tableline
$(0|2)(1|1) $ &
    $ \hbar \alpha_d \left( 6  I_{124} - 2 g_{4R} I_{103} \right)
     Z_{(0|2)R}$ \\ 
\tableline
$(0|2)4 $ & $\hbar \alpha_d  I_{002}  Z_{(0|2)R}$ \\
\tableline
\tableline
$(1|1)0$, $(1|1)1$&  $0 $\\
\tableline
$(1|1)2$ &$\hbar \alpha_d
     \left( 
      3d Z_R I_{024} - 16Z^2_R I_{125}
        \right)/d$ \\
\tableline
$(1|1)(0|1)$  &$ 0$ \\
\tableline
$(1|1)3$ &$ \hbar \alpha_d 
     \left( 8Z_R^2  I_{114} - 2dZ_R I_{013}
             \right)/d$\\
\tableline
$(1|1)(0|2)$ &$0$\\ 
\tableline
$(1|1)(1|1)$  & $1 +  \hbar \alpha_d 
     \left\lbrack
         - 16Z_R^2 I_{225}
      +   \left( 8 +3d \right) Z_R  I_{124} - dI_{023}
             \right\rbrack/d $ \\ 
\tableline
$(1|1)4$  &$ 0$ \\
\tableline
\tableline
$40$, $41$ & $0$ \\
\tableline
$42$ &$ 6 \hbar \alpha_d \left(
     - g_{4R}^2 I_{003} + 6 g_{4R} I_{024} -4 I_{045} \right)$ \\
\tableline
$4(0|1)$  &  $ 0$ \\
\tableline
$43$ &  $ 24 \hbar \alpha_d \left( 
        I_{034} - g_{4R} I_{013}  \right)$ \\
\tableline
$4(0|2)$ &$ 0$ \\
\tableline
$4(1|1)$  &$ 6\hbar \alpha_d \left( 
       - g_{4R}^2 I_{103} + 6 g_{4R} I_{124} - 4 I_{145} \right)$ \\
\tableline
$44$  & $1+  
     6\hbar \alpha_d \left(
     g_{4R} I_{002} - 2 I_{023} 
              \right)$ \\
\tableline
\end{tabular}
\end{table}

\end{document}